\documentclass[12pt]{article}

\usepackage{amssymb}
\usepackage[mathscr]{euscript}

\newtheorem{prop}{Proposition}
\newtheorem{rema}{Remark}
\newtheorem{defi}{Definition}
\newtheorem{lemm}{Lemma}
\newtheorem{theo}{Theorem}
\newtheorem{coro}{Corollary}

\newcommand{\bbox}{\normalsize {}%
        \nolinebreak \hfill $\blacksquare$ \medbreak \par}

\newcommand{\inn}{\hspace*{2pt}\raisebox{-1pt}{\rule{6pt}{.3pt}\hspace*
{0pt}\rule{.3pt}{8pt}\hspace*{3pt}}}

\def\<{\langle} \def\>{\rangle}

\title{Finite dimensional Hamiltonian formalism for gauge and field theories}
\author{  
Fr\'ed\'eric H\'ELEIN\footnote{helein@cmla.ens-cachan.fr} \\
 CMLA, ENS de Cachan \\
 61,avenue du Pr\'esident Wilson\\
 94235 Cachan Cedex, France 
\\ \\
Joseph KOUNEIHER\footnote{kouneiher@paris7.jussieu.fr}\\
Universit\'e Diderot-Paris 7 \\
   case 7064\\
 75005 Paris, France  \\ 
\& \\
CNRS-URA 2052 (C.E.A.)\\
C.E. Saclay
91191 Gif-sur-Yvette Cedex}

\date{May 11, 2000}

\begin{document}
\maketitle
\abstract{We discuss in this paper the canonical structure of classical field theory in finite dimensions within the
{\it{pataplectic}} Hamiltonian formulation, where we put forward the role of Legendre correspondance. We define the generalized
Poisson $\mathfrak{p}$-brackets which are the analogues of the Poisson bracket on forms. We formulate the equations of motion
of forms in terms of $\mathfrak{p}$-brackets.  As illustration of our formalism we present three examples: the interacting scalar
fields, conformal string theory and the electromagnetic field.} 

\section{Introduction}

In the standard Hamiltonian formulation of classical point particle mechanics, the phase space of a system with $N$ degrees of
freedom is a $2N$ dimensional manifold, which represents the space of possible positions and momenta of the system. In field theories
the objects under consideration typically have an infinite number of degrees of freedom. Thus we need to employ infinite dimensional
manifolds to model the dynamical possibilities for such objects. This requires a generalization of the familiar theory of finite
dimensional manifolds, and so we are  motivated to stipulate that an infinite dimensional manifold is a manifold modeled on an
infinite dimensional Banach space.

Infinite dimensional manifolds do, of course, differ in many significant ways from their finite dimensional counterparts. No infinite
dimensional manifold is locally compact, for instance, although every finite dimensional is. Furthermore, the fact that the tangent
spaces are infinite dimensional leads to some complications which are not present in the finite dimensional case. Two are worth
mentioning here:\\
1) In the finite dimensional case, a linear map $T: V \rightarrow V$ is one to one iff it is onto; in the infinite dimensional case
a linear operator on $V$ can be one to one but not onto
\\
2) Similarly, in the finite dimensional case, all linear operators are continuous maps from $V$ to $V$; in the infinite dimensional
case, the continuous linear operators, i.e. the bounded operators, often have as their domain of definition a dense (proper)
subset of $V$.\\ 

A crucial step in the formulation of Hamiltonian mechanics is the construction of the Poisson bracket between a pair of physical
observables. This is obtained from the natural symplectic structure on $T^{\star}M$ (where $M$ is the configuration space of the
physical system). In this phase space approache to classical mechanics, the dynamical evolution from an initial point
$x_{O} \in T^{\star}M$ is the solution to Hamilton's first order differential equations. 
Geometrically, dynamical paths in phase space can be identified with the flow lines of a special vector field $\xi_{H}$ on
$T^{\star}M$ associated with the Hamiltonian fonction $H$.
Those dynamical equations imply the time rate of change of any physical observable
$f \in C^{\infty}(T^{\star}M, \Bbb {R})$, precisely through the Poisson bracket of $f$
with $H$ which is defined thanks to a Hamiltonian vector field $\xi_{f}$ on $T^{\star}M$
associated with $f$. There are couple of potential difficulties here. If $T^{\star}M$ is infinite dimensional, then some
perfectly good functions may not have a Hamiltonian vector field (this problem does not arise in the finite dimensional case).
Even when $\xi_{f}$ exists, its integral curves may be incomplete (i.e., the vector field is only locally defined).

Further, in this viewpoint space and time are treated asymmetrically, therefore we do not have a covariant scheme.

In order to avoid these difficulties, an alternative approach is to construct covariant canonical formulations
of (finite dimensional) field theories which treat the space and time in equal footing (symetrically).
Remark that there is a  whole variety of such theories and interestingly enough they offer a generalization of the Hamilton canonical
equations of motion to field theory, see for instance \cite{De Donder}, \cite{Caratha1}, \cite{Weyl1}, \cite{Rund}. Further
details can be found in \cite{Kastrup1,Gotay1,Gotay2,Gotay3,Rund2,Helein}, and
\cite{Dedecker1,Dedecker2,Gawcedzki1,Kijowski1,Goldschmidt1}. One point there is that the observable quantities
are not represented by (generalized) functions on a phase space, but rather by $n-1$-forms, whose integrals on
Cauchy hypersurfaces give back the usual observables. But many of those approaches share a characteristic, which is
an obstacle to the development of a field quantization, the lack of an appropriate generalization of the Poisson bracket.
And even if a Poisson bracket was proposed, the related construction was too restrictive and not appropriate for representing
the generalized Hamiltonian field equations in Poisson bracket formulation.

More recently, a definition of the Poisson brackets on a subclass of forms and the equations of motion of forms from
De Donder-Weyl point
of view\footnote{In this approach  we associate to the generalized coordinates (the field variables) $u^i$ a set of
$n$ momentum-like  variables (which are defined from the Lagrangian as  the  conjugate  momenta associated with  each
space-time - here $\alpha =1,...,n$ is the space-time index - derivative of the field),
$p^{\alpha}_i:=\frac{\partial L}{\partial (\partial_{\alpha}u^i)}$,  and we have the  Legendre transform:
$\partial_\alpha u^i \rightarrow p^{\alpha}_i$, $$L(u^i, \partial_\alpha u^i, x^\alpha)
\rightarrow H(u^i,p^{\alpha}_i,x^\alpha):= p^{\alpha}_i \partial_\alpha u^i - L$$ So the phase space is replaced by
a finite dimensional space.} was given for review see \cite{kanatchikov1, Kanatchikov2}.
The main point is to derive the Hamiltonian fields equations from the Poincar\'e-Cartan $n$-form and its
differential, called there {\em polysymplectic form} using
``vertical multivector fields'' (which generalize the Hamiltonian vector fields in mechanics). Constructions
of brackets can be done using also the polysymplectic form, but a correct expression of
the dynamics of these forms requires a decomposition of forms and multivectors along
``vertical'' and ``horizontal'' components.
This decomposition, however, essentially implies a triviality of the ``extended
polymomentum phase space'' as a bundle over the space-time manifold. Moreover we notice that in those works 
although a natural link between Poisson brackets and dynamics exists for $n-1$-forms, in the case of forms of arbitrary degrees the
link is not clear\footnote{ This generalization of the Poisson bracket formulation of the equations of motion to forms of arbitrary
degree requires a certain extension. Namely, by adding horizontal forms of degree $n$ and the vertical-vector valued horizontal
one forms (objects of formal degree zero) associated with $n$-forms. This extension, call for a generalization of Lie,
Schouten-Nijenhuis and Fr\"{o}licher-Nijenhuis brackets.}. This affect the posssibility of a precise formulation, for example,
(of the dynamics) of Maxwell's electrodynamics. In addition, we don't have a representation of the energy-momentum
tensor.    \\ 

In this paper we exhibit a general construction of a {\it{universal}} Hamiltonian formalism (which contains all previously
known formalisms, which explains the appelation {\it{universal}}) and define the generalized Poisson $\mathfrak{p}$-brackets,
the analogues of the Poisson bracket on forms. We formulate the equations of motion of forms in terms of those
$\mathfrak{p}$-brackets. The main focus in this construction is on the role of Legendre correspondance, and the hypothesis
concerning the generalized Lengendre condition. We want to emphasize here that in our formalism  there is no need to the
decomposition ``vertical'' and ``horizontal'' parts thanks to the use of Equation $(17)$ which is much more enlightened
than Equation $(15)$. This implies Theorems 2 and 3. On the other hand the energy-momentum tensor is clearly represented and the
Hamiltonian formulation of Maxwell's electrodynamics, for instance, is properly given.

In section (2) we establish the Hamiltonian formalism: the Euler-Lagrange equations, Legendre's correspondance
(and the generalized Legendre condition) and Hamilton's equations. In fact, we recover the Hamilton's equations using
three different approaches: a) as necessary and sufficient conditions for the existence of a critical point
$u: {\cal{X}}\rightarrow {\cal{Y}}$, b) by contracting the {\it{pataplectic}} form $\Omega$ with any $n$-vector
$X\in \Lambda ^nT_{(q,p)}{\cal M}$ where ${\cal M} = \Lambda^nT^{\star}({\cal X}\times {\cal Y})$ and for any
$(q,p)\in {\cal M}$ and finally c) by variational formulation  i.e. as the Euler-Lagrange equations of some simple functional.
In section (3) we review the usual approach to quantum field theory from the  standard canonical viewpoint and  {\it{pataplectic}}
geometry point of view where we express the various brackets using an analogue of the Poisson brackets, the Poisson
$\mathfrak{p}$-brackets, defined on $n-1$-forms. In subsection (3.4) we give a dynamical formulation for the $n-1$-forms in terms
of $\mathfrak{p}$-brackets with the $n$-form ${\cal H}\omega $.

In section (4), and after introducing the internal and external $\mathfrak{p}$-brackets, we generalize the  definition of
$\mathfrak{p}$-bracket
on $n-1$-forms to a class of forms of an arbitrary degrees $0 \leq p \leq n$ using anticommuting (Grassmann) variables
$\tau_{1}$ ...$\tau_{n}$  which behave under change of coordinates like
${\partial \over \partial x^1},\dots ,{\partial \over \partial x^n}$. We should add that those anticommuting variables
do not appear in the expression of the dynamics of forms of arbitrary degrees\footnote{so the role of these anticommuting
variables is similar to the role of ghosts in the quantization of gauge invariant systems}.
Notice also that the generalized Poisson bracket which is obtained here differs from the one
proposed in \cite{kanatchikov1,Kanatchikov2} for forms of degree lower than $p-1$.
In particular ``admissible'' $p$-forms are
composed basically of ``position'' observables unless we have some gauge symmetry and constraints then we can represent some
``momentum'' observable by a $p$-forms with $p \leq n$ (in section (5.3) we study the electromagnetic field which is an instance
of such a situation).  Finally in section (5) we present three examples: the interacting scalar fields, conformal string theory
and the electromagnetic field.

\section{Construction of the Hamiltonian formalism}
In this section we show how to build a universal Hamiltonian formalism for a $\sigma$-model variational problem involving a
Lagrangian functional depending on first derivatives. We derive it through a universal Legendre correspondance.

\subsection{Notations}
Let ${\cal X}$ and ${\cal Y}$ be two differentiable manifolds. ${\cal X}$ plays the role of the space-time
manifold and ${\cal Y}$ the target manifold. We fix some volume form $\omega$ on ${\cal X}$, this
volume form may be chosen according to the variational problem that we want to study (for instance if
we look at the Klein-Gordon functional on some pseudo-Riemannian manifold, we choose $\omega$ to be the Riemannian volume),
but in more general situation, with less symmetries we just choose some arbitrary volume form.
We set $n=\hbox{dim}{\cal X}$ and $k=\hbox{dim}{\cal Y}$. We
denote $\{ x^1,...,x^n\}$ local coordinates on ${\cal X}$ and
$\{ y^1,...,y^k\}$ local coordinates on ${\cal Y}$.
For simplicity we shall assume that the coordinates $x^{\alpha}$
are always chosen such that $dx^1\wedge ...\wedge dx^n =\omega$,
through it is not essential. Then on the product
${\cal X}\times {\cal Y}$ we denote $\{ q^1,...,q^{n+k}\}$ local coordinates in such a
way that

$$\begin{array}{cl}
q^{\mu} = x^{\mu} = x^{\alpha} & \hbox{ if } 1\leq \mu=\alpha \leq n\\
q^{\mu} = y^{\mu-n} = y^i & \hbox{ if } 1\leq \mu-n =i \leq k.
\end{array}$$
Generally we shall denote the
indices running from 1 to $n$ by $\alpha$, $\beta$,... , the indices
between 1 and $k$ by $i$, $j$, ... and the indices between 1 and $n+k$ by
$\mu$, $\nu$,...
To any map $u:{\cal X}\longrightarrow {\cal Y}$ we may associate the map

$$\begin{array}{cccc}
U: & {\cal X} & \longrightarrow & {\cal X}\times {\cal Y}\\
 & x & \longmapsto & (x,u(x))
\end{array}$$
whose image is the graph of $u$, $\{(x,u(x))/x\in {\cal X}\}$.
We also associate to $u$ the bundle $u^{\star}T{\cal Y}\otimes T^{\star}{\cal X}$ over ${\cal X}$.
This bundle is naturally equipped with the coordinates $(x^{\alpha})_{1\leq \alpha \leq n}$
(for ${\cal X}$) and $(v^i_{\alpha})_{1\leq i\leq k;1\leq \alpha \leq n}$, such that a
point $(x,v)\in u^{\star}T{\cal Y}\otimes T^{\star}{\cal X}$ is represented by

$$v = \sum_{\alpha=1}^n\sum_{i=1}^kv^i_{\alpha}{\partial \over \partial y^i}\otimes dx^{\alpha}.$$
We can think $u^{\star}T{\cal Y}\otimes T^{\star}{\cal X}$ as a subset
of $T{\cal Y}\otimes T^{\star}{\cal X}:=\{(x,y,v)/(x,y)\in {\cal X}\times {\cal Y},
v\in T_y{\cal Y}\otimes T^{\star}_x{\cal X}\}$ by the inclusion map
$(x,v)\longmapsto (x,u(x),v)$.

The differential of $u$, $du$ is a section of the bundle $u^{\star}T{\cal Y}\otimes T^{\star}{\cal X}$ over ${\cal X}$. 
Hence the coordinates for $du$ are simply $v^i_{\alpha} = {\partial u^i\over \partial x^{\alpha}}$.
Notice that $u^{\star}T{\cal Y}\otimes T^{\star}{\cal X}$ is a kind of analog of of the tangent bundle
$T{\cal Y}$ to a configuration space ${\cal Y}$ in classical particle mechanics.\\

It turns out to be more convenient to consider $\Lambda^nT({\cal X}\times {\cal Y})$ the analog of $T(\Bbb{R}\times {\cal Y})$, the tangent
bundle to a space-time, or rather $S\Lambda^nT({\cal X}\times {\cal Y})$,  the  submanifold of $\Lambda^nT({\cal X}\times {\cal Y})$, as  the analog of the subset
$ST(\Bbb{R}\times {\cal Y}):=\{(t,x;\xi^0,\vec{\xi})\in T(\Bbb{R}\times {\cal Y})/dt(\xi^0,\vec{\xi})=1\}$,
which is diffeomorphic to $\Bbb{R}\times T{\cal Y}$ by the map $(t,x,\xi)\longmapsto (t,x,\vec{\xi})$, and where:

$$S\Lambda^nT({\cal X}\times {\cal Y}):= \{ (q,z)\in  \Lambda^nT({\cal X}\times {\cal Y})/
z = z_1\wedge ...\wedge z_n, z_1,...,z_n\in T_q({\cal X}\times {\cal Y}), \omega(z_1,...,z_n)=1\}.$$

For any $(x,y)\in {\cal X}\times {\cal Y}$, the fiber $S\Lambda^nT_{(x,y)}({\cal X}\times
{\cal Y})$
can be identified with $T_y{\cal Y}\otimes T^{\star}_x{\cal X}$ by the diffeomorphism

\begin {equation}\label{identify}
\begin{array}{ccc}
T_y{\cal Y}\otimes T^{\star}_x{\cal X} & \longrightarrow &
S\Lambda^nT_{(x,y)}({\cal X}\times {\cal Y})\\
v= \sum_{\alpha=1}^n\sum_{i=1}^kv^i_{\alpha}{\partial \over \partial y^i}\otimes dx^{\alpha}
& \longmapsto &
z = z_1\wedge ...\wedge z_n,
\end{array}
\end{equation}
where for all $1\leq \beta\leq n$, $z_{\beta} = {\partial \over \partial x^{\alpha}} +
\sum_{i=1}^kv^i_{\alpha}{\partial \over \partial y^i}$.
We denote by $(z^{\mu}_{\alpha})_{1\leq \mu\leq n+k;1\leq \alpha \leq n}$ the coordinates of $z_{\alpha}$, so that
$z_{\beta} =\sum_{\mu=1}^{n+k}z^{\mu}_{\alpha}{\partial \over \partial q^{\mu}}$ (or
$z^{\beta}_{\alpha} = \delta^{\beta}_{\alpha}$ for $1\leq \beta\leq n$ and
$z^{\mu+i}_{\alpha} = v^i_{\alpha}$ for $1\leq i\leq k$). This induces an identification
$T{\cal Y}\otimes T^{\star}{\cal X}\simeq S\Lambda^nT({\cal X}\times {\cal Y})$.

Thus coordinates $(x^{\alpha},y^i,v^i_{\alpha})$ (or equivalentely $(x^{\alpha},y^i,z^{\mu}_{\alpha})$) can be thought as coordinate on
$T{\cal Y}\otimes T^{\star}{\cal X}$ or $S\Lambda^nT({\cal X}\times {\cal Y})$.

Given a Lagrangian function $L:T{\cal Y}\otimes T^{\star}{\cal X}\longmapsto \Bbb{R}$, we define the
functional

$${\cal L}[u]:= \int_{\cal X}L(x,u(x),du(x))dx.$$

\subsection{The Euler-Lagrange equations}
The critical points of the action are the maps $u:{\cal X}\longrightarrow {\cal Y}$ which are solutions
of the system ofEuler-Lagrange equations 

\begin{equation}\label{EL}
{\partial \over \partial x^{\alpha}}
\left( {\partial L\over \partial v^i_{\alpha}}(x,u(x),du(x))\right)
= {\partial L\over \partial y^i}(x,u(x),du(x)).
\end{equation}

This equation implies also other equations involving the {\em stress-energy} tensor
associated to $u:{\cal X}\longrightarrow {\cal Y}$:

$$S^{\alpha}_{\beta}(x):= \delta^{\alpha}_{\beta}L(x,u(x),du(x)) -
{\partial L\over \partial v^i_{\alpha}}(x,u(x),du(x))
{\partial u^i\over \partial x^{\beta}}(x).$$
Indeed for any $u$,

$$\begin{array}{ccl}
\displaystyle {\partial S^{\alpha}_{\beta}\over \partial x^{\alpha}}(x) & = &
\displaystyle \delta^{\alpha}_{\beta}\left(
{\partial L\over \partial x^{\alpha}}(x,u,du) +
{\partial L\over \partial y^i}(x,u,du){\partial u^i\over \partial x^{\alpha}}(x) +
{\partial L\over \partial v^i_{\gamma}}(x,u,du)
{\partial ^2 u^i\over \partial x^{\alpha}\partial x^{\gamma}}(x)\right) \\
 & & \displaystyle - {\partial \over \partial x^{\alpha}}
\left( {\partial L\over \partial v^i_{\alpha}}(x,u,du)\right)
{\partial u^i\over \partial x^{\beta}}(x)
- {\partial L\over \partial v^i_{\alpha}}(x,u,du)
{\partial ^2 u^i\over \partial x^{\alpha}\partial x^{\beta}}(x)\\
 & = & \displaystyle {\partial L\over \partial x^{\beta}}(x,u,du)
- \left[ {\partial \over \partial x^{\alpha}}
\left( {\partial L\over \partial v^i_{\alpha}}(x,u(x),du(x))\right)
- {\partial L\over \partial y^i}(x,u(x),du(x))\right]
{\partial u^i\over \partial x^{\beta}}(x).
\end{array}$$
Thus we conclude that if $u$ is a solution of (\ref{EL}), then

\begin{equation}\label{stress}
{\partial S^{\alpha}_{\beta}\over \partial x^{\alpha}}(x) =
{\partial L\over \partial x^{\beta}}(x,u,du).
\end{equation}
It follows that if $L$ does not depend on $x$, then $S^{\alpha}_{\beta}$ is divergence-free
for all solutions of (\ref{EL}), a property which can be predicted by Noether's
theorem.

\subsection{The Legendre correspondance}
Let ${\cal M}:= \Lambda^nT^{\star}({\cal X}\times {\cal Y})$. Every point $(q,p)\in {\cal M}$ has coordinates
$q^{\mu}$ and $p_{\mu_1...\mu_n}$ such that $p_{\mu_1...\mu_n}$ is completely antisymmetric
in $(\mu_1,...,\mu_n)$ and

$$p = \sum_{\mu_1<...<\mu_n}p_{\mu_1...\mu_n}dq^{\mu_1}\wedge ...\wedge dq^{\mu_n}.$$
We shall define a Legendre correspondance 

$$\begin{array}{ccc}
S\Lambda^nT({\cal X}\times {\cal Y})\times \Bbb{R} & \longleftrightarrow & {\cal M}= \Lambda^nT^{\star}({\cal X}\times {\cal Y})\\
(q,v,w) & \longleftrightarrow & (q,p),
\end{array}$$
where $w\in \Bbb{R}$ is some extra parameter (its signification is not clear for the moment, $w$
is related to the possibility of fixing arbitrarely the value of some Hamiltonian). Notice that we do not name it
a transform, like in the classical theory but a correspondance, since generally there will be many possible
values of $(q,p)$ corresponding to a single value of $(q,v,w)$. But we expect that in generic situations,
there corresponds a unique $(q,v,w)$ to some $(q,p)$. This correspondance is generated by the
function

$$\begin{array}{cccc}
W: & S\Lambda^nT({\cal X}\times {\cal Y})\times {\cal M} & \longrightarrow & \Bbb{R}\\
 & (q,v,p) & \longmapsto & \langle p,v\rangle -L(q,v),
\end{array}$$
where

$$\langle p,v\rangle \simeq \langle p,z\rangle :=
\langle p,z_1\wedge ...\wedge z_n\rangle =
\sum_{\mu_1,\dots ,\mu_n}p_{\mu_1...\mu_n}z_1^{\mu_1}\dots z_n^{\mu_n}.$$

\begin{defi}
We write that $(q,v,w) \longleftrightarrow (q,p)$ if and only if
\begin{equation}\label{lw}
L(q,v) + w = \langle p,v\rangle\quad \hbox{or}\quad W(q,v,p) = w
\end{equation}
and
\begin{equation}\label{dlp}
{\partial L\over \partial v^i_{\alpha}}(q,v) =
{\partial \langle p,v\rangle \over \partial  v^i_{\alpha}} =
\left\langle p,z_1\wedge \dots \wedge z_{\alpha-1}\wedge {\partial \over \partial y^i}\wedge z_{\alpha+1}\wedge
\dots \wedge z_n\right\rangle \quad \hbox{or}\quad {\partial W\over \partial v^i_{\alpha}}(q,v,p) = 0.
\end{equation}
\end{defi}
Notice that for any $(q,v,w)\in S\Lambda^nT({\cal X}\times {\cal Y})\times \Bbb{R}$ there exist
$(q,p)\in {\cal M}$ such that $(q,v,w) \longleftrightarrow (q,p)$. This will be proven in Subsection 2.6
below. But $(q,p)$ is not unique in general.
In the following we shall need to suppose that the inverse correspondance is well-defined.\\

\noindent {\bf Hypothesis: Generalized Legendre condition} {\em There exists an open subset ${\cal O}\subset {\cal M}$ which
is non empty such that for any $(q,p)\in {\cal O}$ there exists a unique $v\in T_x{\cal Y}\otimes T_y^{\star}{\cal X}$ (or
equivalentely a unique $z\in S\Lambda^nT_q({\cal X}\times {\cal Y})$) which is a critical point of
$v\longmapsto W(q,v,p)$. We denote $v={\cal V}(q,p)$ this unique solution (or $z={\cal Z}(q,p)$). We
assume further that ${\cal V}$ is a smooth function on ${\cal O}$ (or the same for ${\cal Z}$).}\\

We now suppose that this hypothesis is true. Then we can define on ${\cal O}$ the following
Hamiltonian function

$$\begin{array}{cccc}
{\cal H}: & {\cal O} & \longrightarrow & \Bbb{R}\\
 & (q,p) & \longmapsto & \langle p,{\cal V}(q,p)\rangle - L(q,{\cal V}(q,p))
=W(q,{\cal V}(q,p),p).
\end{array}$$
We then remark that (\ref{lw}) is equivalent to $w = {\cal H}(q,p)$.

We now compute the differential of ${\cal H}$. The main point is to exploit the
condition

\begin{equation}\label{defw}
{\partial W\over \partial v^i_{\alpha}}\left( q,{\cal V}(q,p),p\right) =0
\end{equation}
(which defines ${\cal V}$).

$$\begin{array}{ccl}
d{\cal H} & = & \displaystyle \sum_{\mu}{\partial W\over \partial q^{\mu}}
\left( q,{\cal V}(q,p),p\right) dq^{\mu} +
\sum_{\mu,\nu}\sum_{\alpha}{\partial W\over \partial v^{\nu}_{\alpha}}
\left( q,{\cal V}(q,p),p\right) 
{\partial {\cal V}^{\nu}_{\alpha}\over \partial q^{\mu}}dq^{\mu}\\
 & & \displaystyle +
\sum_{\nu,\alpha}\sum_{\mu_1<...<\mu_n}
{\partial W\over \partial v^{\nu}_{\alpha}}\left( q,{\cal V}(q,p),p\right) 
{\partial {\cal V}^{\nu}_{\alpha}\over \partial p_{\mu_1...\mu_n}}dp_{\mu_1...\mu_n}\\
 & & \displaystyle + \sum_{\mu_1<...<\mu_n}
{\partial W\over \partial p_{\mu_1...\mu_n}}\left( q,{\cal V}(q,p),p\right)
dp_{\mu_1...\mu_n}\\
 & = & \displaystyle \sum_{\mu}{\partial W\over \partial q^{\mu}}
\left( q,{\cal V}(q,p),p\right) dq^{\mu} +
\sum_{\mu_1<...<\mu_n}{\partial W\over \partial p_{\mu_1...\mu_n}}
\left( q,{\cal V}(q,p),p\right) dp_{\mu_1...\mu_n}.
\end{array}$$

Now since

$${\partial W\over \partial q^{\mu}}(q,v,p) = - {\partial L\over \partial q^{\mu}}(q,v,p)$$
and

$${\partial W\over \partial p_{\mu_1...\mu_n}}(q,v,p) =
\left| \begin{array}{ccc}
z^{\mu_1}_1 & \dots & z^{\mu_1}_n\\
\vdots & & \vdots \\
z^{\mu_n}_1 & \dots & z^{\mu_n}_n\\
\end{array}\right| ,$$
we get

\begin{equation}\label{dH}
d{\cal H} = - \sum_{\mu}{\partial L\over \partial q^{\mu}}
\left( q,{\cal V}(q,p),p\right) dq^{\mu} +
\sum_{\mu_1<...<\mu_n} {\cal Z}^{\mu_1...\mu_n}_{1...n}(q,p)dp_{\mu_1...\mu_n},
\end{equation}
where

$${\cal Z}^{\mu_1...\mu_n}_{1...n}(q,p) :=
\left| \begin{array}{ccc}
{\cal Z}^{\mu_1}_1(q,p) & \dots & {\cal Z}^{\mu_1}_n(q,p)\\
\vdots & & \vdots \\
{\cal Z}^{\mu_n}_1(q,p) & \dots & {\cal Z}^{\mu_n}_n(q,p)\\
\end{array}\right| $$
are the components of the $n$-vector

$${\cal Z}_1(q,p)\wedge ...\wedge {\cal Z}_n(q,p) =
\sum_{\mu_1<...<\mu_n} {\cal Z}^{\mu_1...\mu_n}_{1...n}(q,p)
{\partial \over \partial q^{\mu_1}}\wedge \dots \wedge 
{\partial \over \partial q^{\mu_n}}.$$

\noindent To conclude let us see how the stress-energy tensor appears in
this Hamiltonian setting. We define the Hamiltonian tensor on ${\cal O}$ to be
$H(q,p) = \sum_{\alpha,\beta}H^{\alpha}_{\beta}(q,p)
{\partial \over \partial x^{\alpha}}\otimes dx^{\beta}$, with

$$H^{\alpha}_{\beta}(q,p) := 
{\partial L\over \partial v^i_{\alpha}}(q,{\cal V}(q,p)){\cal V}^i_{\beta}(q,p)
- \delta^{\alpha}_{\beta}L(q,{\cal V}(q,p)).$$
It is clear that if $(x,u(x),du(x),w)\longleftrightarrow (q,p)$ then

$$H^{\alpha}_{\beta}(q,p) = - S^{\alpha}_{\beta}(x).$$

Let us now compute $H^{\alpha}_{\beta}(q,p)$. We first use (\ref{dlp})\\

\noindent $\displaystyle \sum_i{\partial L\over \partial v^i_{\alpha}}(q,{\cal V}(q,p))
{\cal V}^i_{\beta}(q,p)$
$$\begin{array}{cl}
= & \displaystyle \sum_i
{\partial \langle p,v\rangle \over \partial  v^i_{\alpha}}_{|v={\cal V}(q,p)}
{\cal V}^i_{\beta}(q,p)\\
= & \displaystyle \sum_i
\left\langle p,{\cal Z}_1(q,p)\wedge ...\wedge {\cal Z}_{\alpha-1}(q,p)\wedge
{\partial \over \partial y^i}\wedge {\cal Z}_{\alpha+1}(q,p)\wedge ...
\wedge {\cal Z}_n(q,p)\right\rangle {\cal V}^i_{\beta}(q,p)\\
= & \displaystyle \sum_{\mu}
\left\langle p,{\cal Z}_1(q,p)\wedge ...\wedge {\cal Z}_{\alpha-1}(q,p)\wedge
{\partial \over \partial q^{\mu}}
\wedge {\cal Z}_{\alpha+1}(q,p)\wedge ...
\wedge {\cal Z}_n(q,p)\right\rangle {\cal Z}^{\mu}_{\beta}(q,p)\\
& \displaystyle
- \left\langle p,{\cal Z}_1(q,p)\wedge ...\wedge {\cal Z}_{\alpha-1}(q,p)\wedge
{\partial \over \partial x^{\beta}}\wedge {\cal Z}_{\alpha+1}(q,p)\wedge ...
\wedge {\cal Z}_n(q,p)\right\rangle \\
= & \displaystyle
\left\langle p,{\cal Z}_1(q,p)\wedge ...\wedge {\cal Z}_{\alpha-1}(q,p)\wedge
{\cal Z}_{\beta}(q,p)\wedge {\cal Z}_{\alpha+1}(q,p)\wedge ...
\wedge {\cal Z}_n(q,p)\right\rangle \\
& \displaystyle
- \left\langle p,{\cal Z}_1(q,p)\wedge ...\wedge {\cal Z}_{\alpha-1}(q,p)\wedge
{\partial \over \partial x^{\beta}}\wedge {\cal Z}_{\alpha+1}(q,p)\wedge ..
\wedge {\cal Z}_n(q,p)\right\rangle \\
= & \displaystyle
\delta^{\alpha}_{\beta}\left\langle p,{\cal Z}_1(q,p)\wedge ...
\wedge {\cal Z}_n(q,p)\right\rangle \\
& \displaystyle
- \left\langle p,{\cal Z}_1(q,p)\wedge ...\wedge {\cal Z}_{\alpha-1}(q,p)\wedge
{\partial \over \partial x^{\beta}}\wedge {\cal Z}_{\alpha+1}(q,p)\wedge ...
\wedge {\cal Z}_n(q,p)\right\rangle .
\end{array}$$
Hence since

$$\left\langle p,{\cal Z}_1(q,p)\wedge ...
\wedge {\cal Z}_n(q,p)\right\rangle = {\cal H}(q,p) + L(q,{\cal V}(q,p)),$$

\begin{equation}\label{tensh}
\begin{array}{ccl}
H^{\alpha}_{\beta}(q,p) & = & \displaystyle \delta^{\alpha}_{\beta}{\cal H}(q,p)
- \left\langle p,{\cal Z}_1(q,p)\wedge ...{\cal Z}_{\alpha-1}(q,p)\wedge
{\partial \over \partial x^{\beta}}\wedge {\cal Z}_{\alpha+1}(q,p)...
\wedge {\cal Z}_n(q,p)\right\rangle \\
 & = & \displaystyle \delta^{\alpha}_{\beta}{\cal H}(q,p)
- {\partial \langle p,z\rangle \over \partial  z^{\beta}_{\alpha}}_{|z={\cal Z}(q,p)}.
\end{array}
\end{equation}

\subsection{Hamilton equations}
Let $x\longmapsto (q(x),p(x))$ be some map from ${\cal X}$ to ${\cal O}$. To insure that this map is related to a critical point $u:{\cal X}\longrightarrow {\cal Y}$, we find that the necessary and sufficient conditions split in two parts: \\

\noindent {\bf 1) What are the conditions on $x\longmapsto (q(x),p(x))$ for the
existence of a map $x\longmapsto u(x)$ such that
$(x,u(x),du(x))\longleftrightarrow (q(x),p(x))$ ?}\\

The first obvious condition is $q(x) = (x,u(x)) = U(x)$. The second condition
is that in $T{\cal Y}\otimes T^{\star}{\cal X}$,
$(x,y,v^i_{\alpha}) = (x,y,{\partial u^i\over \partial x^{\alpha}})$ coincides with
$(q(x),{\cal V}^i_{\alpha}(q(x),p(x)))$. If we translate that using (\ref{identify}),
we obtain that in $S\Lambda^nT({\cal X}\times {\cal Y})$,

$${\partial q\over \partial x^1}\wedge \dots \wedge {\partial q\over \partial x^n} =
{\partial U\over \partial x^1}\wedge \dots \wedge {\partial U\over \partial x^n} =
{\cal Z}_1(q(x),p(x))\wedge \dots \wedge {\cal Z}_n(q(x),p(x)).$$
But we found in (\ref{dH}) that the components in the basis
$\left( {\partial \over \partial q^{\mu_1}}
\wedge \dots \wedge {\partial \over \partial q^{\mu_n}}
\right)$ of the right hand side are
${\cal Z}^{\mu_1\dots \mu_n}_{1\dots n}(q(x),p(x)) =
{\partial {\cal H}\over \partial p_{\mu_1\dots \mu_n}}(q(x),p(x))$. Hence
denoting

$${\partial (q^{\mu_1},\dots ,q^{\mu_n})\over \partial (x^1,\dots ,x^n)} :=
\left| \begin{array}{ccc}
{\partial q^{\mu_1}\over \partial x^1} & \dots & {\partial q^{\mu_1}\over \partial x^n}\\
\vdots & & \vdots \\
{\partial q^{\mu_n}\over \partial x^1} & \dots & {\partial q^{\mu_n}\over \partial x^n}
\end{array}\right| ,$$
so that

$${\partial q\over \partial x^1}\wedge \dots \wedge {\partial q\over \partial x^n} =
\sum_{\mu_1<\dots <\mu_n}
{\partial (q^{\mu_1},\dots ,q^{\mu_n})\over \partial (x^1,\dots ,x^n)}
{\partial \over \partial q^{\mu_1}}\wedge \dots \wedge {\partial \over \partial q^{\mu_n}},$$
we obtain the condition

\begin{equation}\label{hamilton1}
{\partial (q^{\mu_1},\dots ,q^{\mu_n})\over \partial (x^1,\dots ,x^n)}(x) =
{\partial {\cal H}\over \partial p_{\mu_1\dots \mu_n}}(q(x),p(x)).
\end{equation}

\noindent {\bf 2) Now what are the conditions on $x\longmapsto (q(x),p(x))$ for
$u$ to be a solution of the Euler-Lagrange equations ?}\\

It amounts to eliminate $u$ in (\ref{EL}) in function of $(q,p)$. For that
purpose we use (\ref{dlp}) to derive\\

\noindent $\displaystyle \sum_{\alpha}{\partial \over \partial x^{\alpha}}
\left( {\partial L\over \partial v^i_{\alpha}}(x,u(x),du(x))\right) $
$$\begin{array}{cl}
= &
\displaystyle \sum_{\alpha}{\partial \over \partial x^{\alpha}}
\left\langle p,{\partial U\over \partial x^1}\wedge \dots \wedge
{\partial U\over \partial x^{\alpha -1}}\wedge
{\partial \over \partial y^i}\wedge
{\partial U\over \partial x^{\alpha +1}}\wedge \dots \wedge
{\partial U\over \partial x^n}\right\rangle \\
= & \displaystyle \sum_{\alpha}
\left\langle {\partial p\over \partial x^{\alpha}},
{\partial U\over \partial x^1}\wedge \dots \wedge
{\partial U\over \partial x^{\alpha -1}}\wedge
{\partial \over \partial y^i}\wedge
{\partial U\over \partial x^{\alpha +1}}\wedge \dots \wedge
{\partial U\over \partial x^n}\right\rangle \\
& + \displaystyle \sum_{\alpha\neq \beta}
\left\langle p,{\partial U\over \partial x^1}\wedge \dots \wedge
{\partial ^2U\over \partial x^{\alpha}\partial x^{\beta}}
\wedge \dots \wedge
{\partial U\over \partial x^{\alpha -1}}\wedge
{\partial \over \partial y^i}\wedge
{\partial U\over \partial x^{\alpha +1}}\wedge \dots \wedge
{\partial U\over \partial x^n}\right\rangle \\
= & \displaystyle \sum_{\alpha}
\left\langle {\partial p\over \partial x^{\alpha}},
{\partial U\over \partial x^1}\wedge \dots \wedge
{\partial U\over \partial x^{\alpha -1}}\wedge
{\partial \over \partial y^i}\wedge
{\partial U\over \partial x^{\alpha +1}}\wedge \dots \wedge
{\partial U\over \partial x^n}\right\rangle .
\end{array}$$

On the other hand we know from (\ref{dH}) that
${\partial {\cal H}\over \partial q^i}(q,p) =
- {\partial L\over \partial q^i}(q,{\cal V}(q,p))$. Thus we obtain

\begin{equation}\label{hamilton2}
\sum_{\alpha}
\left\langle {\partial p\over \partial x^{\alpha}},
{\partial q\over \partial x^1}\wedge \dots \wedge
{\partial q\over \partial x^{\alpha -1}}\wedge
{\partial \over \partial y^i}\wedge
{\partial q\over \partial x^{\alpha +1}}\wedge \dots \wedge
{\partial q\over \partial x^n}\right\rangle =
- {\partial {\cal H}\over \partial q^i}(q(x),p(x)).
\end{equation}
The latter equation may be transformed using the relation

\noindent $\displaystyle \sum_{\alpha}
\left\langle {\partial p\over \partial x^{\alpha}},
{\partial q\over \partial x^1}\wedge \dots \wedge
{\partial q\over \partial x^{\alpha -1}}\wedge
{\partial \over \partial y^i}\wedge
{\partial q\over \partial x^{\alpha +1}}\wedge \dots \wedge
{\partial q\over \partial x^n}\right\rangle$
$$\begin{array}{cl}
= & \displaystyle
\sum_{\alpha}
\sum_{\footnotesize \begin{array}{c}\mu_1<\dots <\mu_n\\\mu_{\alpha} = n+i\end{array}}
\left| \begin{array}{ccc}
{\partial q^{\mu_1}\over \partial x^1} & \dots & {\partial q^{\mu_1}\over \partial x^n}\\
\vdots & & \vdots \\
{\partial q^{\mu_{\alpha -1}}\over \partial x^1} & \dots & {\partial q^{\mu_{\alpha -1}}\over \partial x^n}\\
{\partial p_{\mu_1\dots \mu_n}\over \partial x^1} &
\dots & {\partial p_{\mu_1\dots \mu_n}\over \partial x^n}\\
{\partial q^{\mu_{\alpha +1}}\over \partial x^1} & \dots & {\partial q^{\mu_{\alpha +1}}\over \partial x^n}\\
\vdots & & \vdots \\
{\partial q^{\mu_n}\over \partial x^1} & \dots & {\partial q^{\mu_n}\over \partial x^n}
\end{array}\right| \\
 & \\
=  & \displaystyle \sum_{\alpha}
\sum_{\footnotesize \begin{array}{c}\mu_1<\dots <\mu_n\\\mu_{\alpha} = n+i\end{array}}
{\partial (q^{\mu_1},\dots ,q^{\mu_{\alpha -1}},
p_{\mu_1\dots \mu_n}, q^{\mu_{\alpha +1}},\dots ,
q^{\mu_n})\over \partial (x^1,\dots ,x^n)}.
\end{array}$$

We summarize: the necessary and sufficient conditions we were looking for are

\begin{equation}\label{hamilton3}
\begin{array}{c}\displaystyle
{\partial (q^{\mu_1},\dots ,q^{\mu_n})\over \partial (x^1,\dots ,x^n)} =
{\partial {\cal H}\over \partial p_{\mu_1\dots \mu_n}}(q,p)\\
\\
\displaystyle
\sum_{\alpha}
\sum_{\footnotesize \begin{array}{c}\mu_1<\dots <\mu_n\\\mu_{\alpha} = n+i\end{array}}
{\partial (q^{\mu_1},\dots ,q^{\mu_{\alpha -1}},
p_{\mu_1\dots \mu_n}, q^{\mu_{\alpha +1}},\dots ,
q^{\mu_n})\over \partial (x^1,\dots ,x^n)} 
= - {\partial {\cal H}\over \partial y^i}(q,p).
\end{array}
\end{equation}

\noindent{\bf Some further relations}\\

Besides these equations, we have to remark also that equation (\ref{stress}) on the
stress-energy tensor has a counterpart in this formalism. For that purpose we
use equation (\ref{tensh}). Assuming that $(x,u(x),du(x))\longleftrightarrow
(q(x),p(x))$, we have

$$\begin{array}{ccl}
\displaystyle
- {\partial S^{\alpha}_{\beta}\over \partial x^{\alpha}}(x) & = &
\displaystyle
{\partial H^{\alpha}_{\beta}(q(x),p(x))\over \partial x^{\alpha}} \\
 & = &\displaystyle 
{\partial {\cal H}(q(x),p(x))\over \partial x^{\beta}}
- {\partial \over \partial x^{\alpha}}\left\langle
p(x),{\partial q(x)\over \partial x^1}\wedge \dots \wedge
{\partial q(x)\over \partial x^{\alpha -1}}\wedge
{\partial \over \partial x^{\beta}}\wedge
{\partial q(x)\over \partial x^{\alpha +1}}\wedge \dots \wedge
{\partial q(x)\over \partial x^n}\right\rangle \\
 & = & \displaystyle
{\partial {\cal H}(q(x),p(x))\over \partial x^{\beta}}
- \left\langle
{\partial p(x)\over \partial x^{\alpha}},
{\partial q(x)\over \partial x^1}\wedge \dots \wedge
{\partial q(x)\over \partial x^{\alpha -1}}\wedge
{\partial \over \partial x^{\beta}}\wedge
{\partial q(x)\over \partial x^{\alpha +1}}\wedge \dots \wedge
{\partial q(x)\over \partial x^n}\right\rangle .
\end{array}$$
Now assume that $u$ is a critical point, then because of (\ref{stress}) and (\ref{dH}),

$${\partial S^{\alpha}_{\beta}\over \partial x^{\alpha}}(x) =
{\partial L\over \partial x^{\beta}}(x,u(x),du(x)) =
- {\partial {\cal H}\over \partial x^{\beta}}(q(x),p(x)).$$
And we obtain

$$
\left\langle
{\partial p\over \partial x^{\alpha}},
{\partial q\over \partial x^1}\wedge \dots \wedge
{\partial q\over \partial x^{\alpha -1}}\wedge
{\partial \over \partial x^{\beta}}\wedge
{\partial q\over \partial x^{\alpha +1}}\wedge \dots \wedge
{\partial q\over \partial x^n}\right\rangle  
- {\partial \over \partial x^{\beta}}\left( {\cal H}(q,p)\right)
= - {\partial {\cal H}\over \partial x^{\beta}}(q,p)
$$
or equivalentely

\begin{equation}\label{hamilton4}
\sum_{\alpha}
\sum_{\footnotesize \begin{array}{c}\mu_1<\dots <\mu_n\\\mu_{\alpha} = \beta \end{array}}
{\partial (q^{\mu_1},\dots ,q^{\mu_{\alpha -1}},
p_{\mu_1\dots \mu_n}, q^{\mu_{\alpha +1}},\dots ,
q^{\mu_n})\over \partial (x^1,\dots ,x^n)} 
- {\partial \over \partial x^{\beta}}\left( {\cal H}(q,p)\right)
= - {\partial {\cal H}\over \partial x^{\beta}}(q,p).
\end{equation}

\noindent {\bf Conclusion} The Hamilton equations (\ref{hamilton3}) can be
completed by adding (\ref{hamilton4}) (which are actually a consequence of
(\ref{hamilton3})). We thus obtain 

\begin{equation}\label{hamilton5}
\begin{array}{c}\displaystyle
{\partial (q^{\mu_1},\dots ,q^{\mu_n})\over \partial (x^1,\dots ,x^n)} =
{\partial {\cal H}\over \partial p_{\mu_1\dots \mu_n}}(q,p)\\
\\
\displaystyle
\sum_{\alpha}
\sum_{\footnotesize \begin{array}{c}\mu_1<\dots <\mu_n\\\mu_{\alpha} = \nu\end{array}}
{\partial (q^{\mu_1},\dots ,q^{\mu_{\alpha -1}},
p_{\mu_1\dots \mu_n}, q^{\mu_{\alpha +1}},\dots ,
q^{\mu_n})\over \partial (x^1,\dots ,x^n)}
- \sum_{\alpha}\delta^{\alpha}_{\nu}
{\partial \over \partial x^{\alpha}}\left( {\cal H}(q,p)\right)
= - {\partial {\cal H}\over \partial q^{\nu}}(q,p).
\end{array}
\end{equation}

\subsection{The Cartan-Poincar\'e and pataplectic forms
on\\${\cal M} = \Lambda^nT^{\star}({\cal X}\times {\cal Y})$}
Motivated by the previous contruction, we define the Cartan-Poincar\'e form on
$\Lambda^nT^{\star}({\cal X}\times {\cal Y})$ to be

$$\theta := \sum_{\mu_1<\dots <\mu_n}p_{\mu_1\dots \mu_n}dq^{\mu_1}\wedge \dots \wedge dq^{\mu_n}.$$
Its differential is

$$\Omega := \sum_{\mu_1<\dots <\mu_n}dp_{\mu_1\dots \mu_n}\wedge dq^{\mu_1}\wedge \dots \wedge dq^{\mu_n},$$
which we will call the {\em pataplectic form}, a straightforward generalization
of the symplectic form.\\

A first property is that we can express the system of Hamilton's equations (\ref{hamilton5}) in an elegant way using $\Omega$. For any $(q,p)\in {\cal M}$ and any $n$-vector
$X\in \Lambda ^nT_{(q,p)}{\cal M}$
we define $X\inn \Omega \in T^{\star}_{(q,p)}{\cal M}$ as follows.
If $X$ is decomposable, i. e. if there exist $n$ vectors $X_1,\dots ,X_n\in T_{(q,p)}{\cal M}$ such that
$X = X_1\wedge \dots \wedge X_n$, we let

$$X\inn \Omega(V) := \Omega(X_1,\dots ,X_n,V),\quad \forall V\in T_{(q,p)}{\cal M}.$$
We extend this definition to non decomposable $X$ by linearity. Let us analyse $X\inn \Omega$ using
coordinates. Writing $X$ as 

{\small
$$\begin{array}{l}
\displaystyle \sum_{\mu_1<\dots <\mu_n}X^{\mu_1\dots \mu_n}
\partial _{\mu_1}\wedge \dots \wedge \partial _{\mu_n}\\
+\displaystyle \sum_{\footnotesize \begin{array}{c}\mu_1<\dots <\mu_{\alpha-1}<\mu_{\alpha+1}<\dots <\mu_n \\ \nu_1<\dots <\nu_n\end{array}}
X{^{\mu_1\dots \mu_{\alpha-1}}}{_{ \{\nu_1\dots \nu_n\} }} {^{\mu_{\alpha+1}\dots \mu_n}}
\partial _{\mu_1}\wedge \dots \wedge \partial _{\mu_{\alpha-1}}
\wedge \partial ^{\nu_1\dots \nu_n}\wedge \partial _{\mu_{\alpha+1}}
\wedge \dots \wedge \partial _{\mu_n}\\
+\quad \hbox{etc }\dots
\end{array}$$}
with the notations $\partial _{\mu}:= {\partial \over \partial q^{\mu}}$,
$\partial ^{\nu_1\dots \nu_n}:= {\partial \over \partial p_{\nu_1\dots \nu_n}}$, we have

$$X\inn\Omega = (-1)^n\left[
\sum_{\mu_1<\dots <\mu_n}X^{\mu_1\dots \mu_n} dp_{\mu_1\dots \mu_n}
- \sum_{\nu}\sum_{\alpha}
\sum_{\footnotesize \begin{array}{c}\mu_1<\dots <\mu_n\\\mu_{\alpha} = \nu\end{array}}
X{^{\mu_1\dots \mu_{\alpha-1}}} {_{\{\mu_1\dots \mu_n\} }} {^{\mu_{\alpha+1}\dots \mu_n}} dq^{\nu}
\right] .$$
Algebraic similarities with (\ref{hamilton5}) are evident if we replace $X$ by
${\partial (q,p)\over \partial (x^1,\dots ,x^n)}:=
{\partial (q,p)\over \partial x^1}\wedge \dots
\wedge {\partial (q,p)\over \partial x^n}$. In particular we can see easily that the coefficients of $dy^i$ and $dp_{\mu_1\dots \mu_n}$ in
$(-1)^n{\partial (q,p)\over \partial (x^1,\dots ,x^n)}\inn \Omega$
and $d{\cal H}$ coincide if and only if the Hamilton system (\ref{hamilton3}) holds.
Thus we are led to define ${\cal I}$ to be the algebraic ideal in
$\Lambda^{\star}{\cal M}$ spanned by $\{dx^1,\dots ,dx^n\}$ and hence (\ref{hamilton3})
is equivalent to

\begin{equation}\label{hamilton7}
(-1)^n{\partial (q,p)\over \partial (x^1,\dots ,x^n)}\inn \Omega =
d{\cal H} \quad \hbox{mod }{\cal I}.
\end{equation}

\begin{defi}
A $n$-vector $X\in \Lambda^nT_{(q,p)}{\cal M}$ is ${\cal H}$-Hamiltonian if
and only if

\begin{equation}\label{hamilvec1}
(-1)^nX\inn \Omega = d{\cal H}\quad \hbox{mod }{\cal I}.
\end{equation}
\end{defi}
For such an $X$, it is possible to precise the relation between the left and right hand sides of
(\ref{hamilvec1}) in the case where $X$ is decomposable, i. e.
$X = X_1\wedge \dots \wedge X_n$. Notice that (\ref{hamilvec1})
implies in particular $X^{1\dots n} = {\partial {\cal H}\over \partial \epsilon} = 1$ (where $ \epsilon := p_{1\dots n}$ see (\ref{hamilton5})),
which is equivalent to $\omega(X_1,\dots ,X_n)=1$. Hence we may
always assume without loss of generality that the $X_{\alpha}$ are chosen
so that $dx^{\beta}(X_{\alpha}) = \delta^{\beta}_{\alpha}$. Such vectors are unique.

\begin{lemm}
Let $X = X_1\wedge \dots \wedge X_n\in \Lambda^nT_{(q,p)}{\cal M}$ such that
$dx^{\beta}(X_{\alpha}) = \delta^{\beta}_{\alpha}$. Then $X$ is
${\cal H}$-Hamiltonian if and only if one of the two following relations are
satisfied:

\begin{equation}\label{hamilvec2}
(-1)^nX\inn \Omega = d{\cal H} - \sum_{\alpha}d{\cal H}(X_{\alpha})dx^{\alpha}.
\end{equation}
or

\begin{equation}\label{hamilvec3}
X\inn (\Omega -d({\cal H}\omega))= 0.
\end{equation}
\end{lemm}
{\bf Proof} Let us prove first that (\ref{hamilvec1}) implies (\ref{hamilvec2}).
Since for any $\alpha$, $\beta$,
$dx^{\beta}\left( X_{\alpha} - {\partial \over \partial x^{\alpha}}\right) =0$,
equation (\ref{hamilvec1}) implies that for all $\alpha$,

$$(-1)^nX\inn \Omega\left( X_{\alpha} - {\partial \over \partial x^{\alpha}}\right) =
d{\cal H}\left( X_{\alpha} - {\partial \over \partial x^{\alpha}}\right)$$
$$\Longleftrightarrow
(-1)^n\Omega\left( X_1,\dots ,X_n,X_{\alpha} - {\partial \over \partial x^{\alpha}}\right)
= d{\cal H}(X_{\alpha}) - {\partial {\cal H}\over \partial x^{\alpha}}$$
$$\Longleftrightarrow
(-1)^nX\inn \Omega\left( {\partial \over \partial x^{\alpha}}\right) =
{\partial {\cal H}\over \partial x^{\alpha}} - d{\cal H}(X_{\alpha}).$$
This implies

\begin{equation}\label{hamilvec0}
(-1)^n\sum_{\alpha}X\inn \Omega
\left( {\partial \over \partial  x^{\alpha}}\right) dx^{\alpha} = 
\sum_{\alpha}{\partial {\cal H}\over \partial x^{\alpha}}dx^{\alpha}
- \sum_{\alpha}d{\cal H}(X_{\alpha})dx^{\alpha}.
\end{equation}
Now if we rewrite (\ref{hamilvec1}) as

\noindent $\displaystyle (-1)^n\left( \sum_iX\inn \Omega
\left( {\partial \over \partial  y^i}\right) dy^i +
\sum_{\mu_1<\dots  <\mu_n}X\inn \Omega
\left( {\partial \over \partial  p_{\mu_1\dots  \mu_n}}\right) dp_{\mu_1\dots  \mu_n}
\right) =$
$$\sum_i {\partial {\cal H}\over \partial y^i}dy^i +
\sum_{\mu_1<\dots  <\mu_n} {\partial {\cal H}\over \partial p_{\mu_1\dots  \mu_n}}
dp_{\mu_1\dots  \mu_n},$$
and sum with (\ref{hamilvec0}), we obtain exactly (\ref{hamilvec2}).\\

\noindent Now relation (\ref{hamilvec3}) is equivalent to (\ref{hamilvec2}) because
of the following calculation. For all vector $V$ and for any decomposable
$n$-vector $X = X_1\wedge \dots \wedge X_n\in \Lambda^nT_{(q,p)}{\cal M}$ such that
$dx^{\beta}(X_{\alpha}) = \delta^{\beta}_{\alpha}$ (not necessarily
${\cal H}$-Hamiltonian), we have

$$\begin{array}{ccl}
X\inn d({\cal H}\omega) (V) & = & \displaystyle
d{\cal H}\wedge \omega (X_1,\dots ,X_n,V)\\
 & = & \displaystyle
\sum_{\alpha}(-1)^{\alpha-1}d{\cal H}(X_{\alpha})
\omega(X_1,\dots ,X_{\alpha-1},X_{\alpha+1},\dots ,X_n,V)\\
& & \displaystyle
+ (-1)^nd{\cal H}(V)\omega(X_1,\dots ,X_n,V)\\
 & = & \displaystyle
\sum_{\alpha}(-1)^{n-1}d{\cal H}(X_{\alpha})dx^{\alpha}(V)
+ (-1)^nd{\cal H}(V)
\end{array}$$
thus
$$X\inn d({\cal H}\omega) =  (-1)^n\left( d{\cal H} -
\sum_{\alpha}d{\cal H}(X_{\alpha})dx^{\alpha}\right)$$
and (\ref{hamilvec2}) $\Longleftrightarrow$ (\ref{hamilvec3}).
Conversely it is obvious that (\ref{hamilvec2}) and (\ref{hamilvec3}) implies
(\ref{hamilvec1}).
\bbox

As a Corollary of this result we deduce that a reformulation of
(\ref{hamilton7}) is

\begin{equation}\label{hamilton6}
{\partial (q,p)\over \partial (x^1,\dots ,x^n)}\inn \Omega =
{\partial (q,p)\over \partial (x^1,\dots ,x^n)}\inn (d{\cal H}\wedge \omega).
\end{equation}
It is an exercise to check that actually this relation is a direct translation
of (\ref{hamilton5}).

\subsection{A variational formulation of (\ref{hamilton5})}

We shall now prove that equations (\ref{hamilton5}) are the Euler-Lagrange equations of some
simple functional. For that purpose, let $\Gamma$ be an oriented submanifold of dimension $n$
in $\Lambda^nT^{\star}({\cal X}\times {\cal Y})$ such that $\omega_{|\Gamma}>0$ everywhere.
Then we define the functional

$${\cal A}[\Gamma] := \int_{\Gamma}\theta -\lambda{\cal H}(q,p)\omega.$$
Here $\lambda$ is a (real) scalar function defined over $\Gamma$ which plays
the role of a Lagrange multiplier. We now characterise submanifolds $\Gamma$ which are critical points
of ${\cal A}$.\\

\noindent {\bf Variations with respect to $p$}\\
Let $\delta p$ be some infinitesimal variation of $\Gamma$ with compact support. We compute

$$\delta{\cal A}_{\Gamma}(\delta p) = \int_{\Gamma}\delta p_{\mu_1\dots \mu_n}
\left( dq^{\mu_1}\wedge \dots \wedge dq^{\mu_n}
-\lambda {\partial {\cal H}\over \partial p_{\mu_1\dots \mu_n}}\omega \right) .$$
Assuming that this vanishes for all $\delta p$, we obtain

$$\left( dq^{\mu_1}\wedge \dots \wedge dq^{\mu_n}\right) _{|\Gamma} =
\lambda {\partial {\cal H}\over \partial p_{\mu_1\dots \mu_n}}\omega_{|\Gamma}.$$
This relation means that for any orientation preserving parametrization
$(t^1,\dots ,t^n)\longmapsto (q,p)(t^1,\dots ,t^n)$
of $\Gamma$,

$${\partial (q^{\mu_1},\dots ,q^{\mu_n})\over \partial (t^1,\dots ,t^n)}
= \lambda {\partial {\cal H}\over \partial p_{\mu_1\dots \mu_n}}\omega
\left( {\partial q^{\mu_1}\over \partial t^1},\dots ,{\partial q^{\mu_n}\over \partial t^n}\right) .$$
But we remark that because ${\partial {\cal H}\over \partial p_{1\dots n}} = 1$, the above relation for
$(\mu_1,\dots ,\mu_n) = (1,\dots ,n)$ forces $\lambda =1$. Hence

$${\cal A}[\Gamma] = \int _{\Gamma}\theta -{\cal H}(q,p)\omega.$$
Moreover the equation obtained here can be written using the natural parametrization
$(x^1,\dots ,x^n)\longmapsto (x,u(x),p(x))$ (for which
$\omega\left( {\partial \over \partial x^1},\dots ,{\partial \over \partial x^n}\right) =1$)
and then we obtain

$${\partial q\over \partial x^1}\wedge \dots \wedge {\partial q\over \partial x^n} =
{\partial {\cal H}\over \partial p}(q,p),$$
i. e. exactly equation (\ref{hamilton1})
\footnote{Note that this relation actually implies
${\cal A}[\Gamma] = \int_{\cal X}L(x,q,dq)\omega$. Hence, as in the
one-dimensional Hamilton formalism, $\theta - {\cal H}\omega$ plays the role of
the Lagrangian density.}.\\

\noindent {\bf Variations with respect to $q$}\\
Now $\delta q$ is some infinitesimal variation of $\Gamma$ with compact support. And we have

$$\delta{\cal A}_{\Gamma}(\delta q) = \int_{\Gamma}\sum_{\mu_1<\dots <\mu_n}\sum_{\alpha}
p_{\mu_1\dots \mu_n}dq^{\mu_1}\wedge \dots \wedge d(\delta q^{\mu_{\alpha}})\wedge \dots  \wedge dq^{\mu_n}
- \sum_{\mu}{\partial {\cal H}\over \partial q^{\mu}}\delta q^{\mu}\omega - {\cal H}(q,p)\delta\omega.$$
We pay  special attention to $\delta\omega$:

$$\delta\omega = d(\delta x^1)\wedge \dots \wedge dx^n + \dots + dx^1\wedge \dots \wedge d(\delta x^n).$$
Hence

$$\begin{array}{ccl}
\displaystyle \int_{\Gamma}{\cal H}(q,p)\delta\omega & = & \displaystyle - \int_{\Gamma}
\delta x^1\left( d({\cal H}(q,p))\wedge \dots \wedge dx^n\right) + \dots +
\delta x^n\left( dx^1\wedge \dots \wedge d({\cal H}(q,p))\right) \\
& = & \displaystyle - \sum _{\alpha}\delta x^{\alpha}{\partial \over \partial x^{\alpha}}\left( {\cal H}(q,p)\right) \omega.
\end{array}$$
Thus after integrations by parts, we obtain

$$\begin{array}{ccl}
\displaystyle \delta{\cal A}_{\Gamma}(\delta q) & = & \displaystyle
\int_{\Gamma} - \sum_{\mu_1<\dots <\mu_n}\sum_{\alpha}\delta q^{\mu_{\alpha}}
dq^{\mu_1}\wedge \dots \wedge dq^{\mu_{\alpha -1}}\wedge dp_{\mu_1\dots \mu_n}\wedge dq^{\mu_{\alpha +1}}
\wedge \dots  \wedge dq^{\mu_n}\\
& & \displaystyle - \sum_{\mu}{\partial {\cal H}\over \partial q^{\mu}}\delta q^{\mu}\omega
+\sum _{\alpha}\delta x^{\alpha}{\partial \over \partial x^{\alpha}}\left( {\cal H}(q,p)\right) \omega.
\end{array}$$
And this vanishes if and only if

$$\sum_{\alpha}\sum_{\begin{array}{c}\mu_1<\dots <\mu_n\\\mu_{\alpha} = \nu\end{array}}
dq^{\mu_1}\wedge \dots \wedge dq^{\mu_{\alpha -1}}\wedge dp_{\mu_1\dots \mu_n}\wedge dq^{\mu_{\alpha +1}}
\wedge \dots  \wedge dq^{\mu_n} -
\sum _{\alpha}\delta^{\alpha}_{\nu}{\partial \over \partial x^{\alpha}}\left( {\cal H}(q,p)\right) \omega=
- {\partial {\cal H}\over \partial q^{\nu}}\omega.$$
Again by choosing the parametrization $(x^1,\dots ,x^n)\longmapsto (x,u(x),p(x))$, this relation is
easily seen to be equivalent to (\ref{hamilton2}) and (\ref{hamilton4}).

By the same token we have proven that if we look to critical points of the functional
$\int_{\Gamma}\theta$ with the constraint ${\cal H}(q,p) = h$, for some constant $h$, then the Lagrange
multiplier is 1 and they satisfy the same equations.

\begin{theo}
Let $\Gamma$ be an oriented submanifold of dimension $n$
in $\Lambda^nT^{\star}({\cal X}\times {\cal Y})$ such that $\Omega_{|\Gamma}>0$ everywhere.
Then the three following assertions are equivalent
\begin{itemize}
\item $\Gamma$ is the graph of a solution of the generalized Hamilton equations (\ref{hamilton5})
\item $\Gamma$ is a critical point of the functional
$\int_{\Gamma}\theta - {\cal H}(q,p)\omega$
\item $\Gamma$ is a critical point of the functional $\int_{\Gamma}\theta$ under the constraint
that ${\cal H}(q,p)$ is constant.
\end{itemize}
\end{theo}

\subsection{Some particular cases}
By restricting the variables $(q,p)$ to lie in some submanifold of
${\cal M}=\Lambda^nT^{\star}({\cal X}\times {\cal Y})$,
the Legendre correspondance becomes in some situations a true map.\\

\noindent {\bf a)} We assume that all components $p_{\mu_1,\dots \mu_n}$ vanishes excepted for

$$p_{1\dots n} =: \epsilon \quad \hbox{and}\quad p_{1\dots (\alpha-1)(n+i)(\alpha+1)\dots n} =: p^{\alpha}_i$$
and all obvious permutations in the indices. This defines a submanifold ${\cal M}_{\hbox {\tiny Weyl}}$ of
${\cal M}$. It means that

$$\theta_{|{\cal M}_{\hbox {\tiny Weyl}}}= \epsilon\; dx^1\wedge \dots \wedge dx^n +
\sum_{\alpha}\sum_ip^{\alpha}_idx^1\wedge \dots \wedge dx^{\alpha-1}\wedge dy^i
\wedge dx^{\alpha+1}\wedge \dots \wedge dx^n.$$
Then for any $(q,p)\in {\cal M}_{\hbox {\tiny Weyl}}$,
$\langle p,z_1\wedge \dots \wedge z_n\rangle = \epsilon + \sum_{\alpha}\sum_ip^{\alpha}_iv^i_{\alpha}$,
$W(q,v,p) = \epsilon + \sum_{\alpha}\sum_ip^{\alpha}_iv^i_{\alpha} - L(q,v)$. Hence the relation
(\ref{dlp}) ${\partial W\over \partial v^i_{\alpha}}(q,v)=0$ is equivalent to

$$p^{\alpha}_i = {\partial L\over \partial v^i_{\alpha}}(q,v)
\Longleftrightarrow v^i_{\alpha} = {\cal V}^i_{\alpha}(q,p) .$$
The relation (\ref{lw}) $W(q,v,p) = w$ gives

$$\epsilon = w + L(q,v) - \sum_{\alpha}\sum_i{\partial L\over \partial v^i_{\alpha}}(q,v)v^i_{\alpha}.$$
Last we have that
${\cal H}(q,p) = \epsilon + \sum_{\alpha}\sum_ip^{\alpha}_i{\cal V}^i_{\alpha}(q,p) - L(q,{\cal V}(q,p))$.

This example shows that  for any
$(q,v,w)\in S\Lambda^nT({\cal X}\times {\cal Y})\times \Bbb{R}$,
there exist $(q,p)\in {\cal M}$ such that $(q,v,w)\longleftrightarrow (q,p)$ and
this $(q,p)$ is unique
if it is chosen in ${\cal M}_{\hbox {\tiny Weyl}}$\\

To summarize, we recover the Weyl theory (see \cite{Rund,Helein}).
As an exercize, the reader can check that
in this situation, equations (\ref{hamilton3}) are equivalent to

\begin{equation}\label{weyl}
{\partial y^i\over \partial x^{\alpha}} = {\partial {\cal H}\over \partial p^{\alpha}_i},\quad
\sum_{\alpha}{\partial p^{\alpha}_i\over \partial x^{\alpha}} = - {\partial {\cal H}\over \partial y^i}.
\end{equation}

\noindent {\bf b)}  We assume that $(q,p)$ are such that there exist coefficients $\left( \pi^{\alpha}_{\mu}\right)_{\alpha,\mu}$
with

$$p_{\mu_1\dots \mu_n} = \left| \begin{array}{ccc}
\pi^1_{\mu_1} & \dots & \pi^1_{\mu_n}\\
\vdots & & \vdots \\
\pi^n_{\mu_1} & \dots & \pi^n_{\mu_n}\\
\end{array}\right| .$$
This constraint defines a submanifold ${\cal M}_{\hbox {\tiny Carath\'eodory}}$ of ${\cal M}$. Then

$$\theta_{|{\cal M}_{\hbox {\tiny Carath\'eodory}}} = \left( \sum_{\mu_1}\pi^1_{\mu_1}dq^{\mu_1}\right) \wedge \dots \wedge
\left( \sum_{\mu_n}\pi^1_{\mu_n}dq^{\mu_n}\right) .$$
Then it is an exercise to see that, by choosing $w=0$, it leads
to the formalism developped in \cite{Rund} and \cite{Helein}
associated to the Carath\'eodory
theory of equivalent integrals. However it is not clear in general whether it is
possible to perform the Legendre transform in this setting by being able
to fix arbitrarirely the value of $w$. It is so if we do not impose a 
condition on $w$.

\section{Comparison with the usual Hamiltonian formalism for quantum fields theory}
\subsection{Reminder of the usual approach to quantum field theory}
Here we compare the preceeding construction with the classical approach to quantum field theory
by  so-called canonical quantization. We shall first explore it in the case where ${\cal X}$ is the
Minkowski space $\Bbb{R}\times \Bbb{R}^{n-1}$ and $y=\phi$ is a real scalar field. Hence ${\cal Y}=\Bbb{R}$.
Our functional is

$${\cal L}[\phi] := \int_{\Bbb{R}\times \Bbb{R}^{n-1}}L(x,\phi,d\phi)dx.$$
For simplicity, we may keep in mind the following example of Lagrangian:

$$\int_{\Bbb{R}\times \Bbb{R}^{n-1}}L(x,\phi,d\phi)dx =
\int_{\Bbb{R}\times \Bbb{R}^{n-1}}\left( {1\over 2}\left( {\partial \phi\over \partial x^0}\right) ^2
- {1\over 2}\sum_{\alpha=1}^{n-1}\left( {\partial \phi\over \partial x^{\alpha}}\right) ^2 - V(\phi)\right) dx^0d\vec{x},$$
where we denote $\vec{x} = (x^{\alpha})_{1\leq \alpha\leq n-1}$. We shall also denote $t=x^0$.

The usual approach consists in selecting a global time coordinate $t$ as we already
implicitely assumed here. Then for each time the instantaneous state of the field is
seen as a point in the infinite dimensional ``manifold''
$\mathfrak{F}:= \{\Phi:\Bbb{R}^{n-1}\longrightarrow \Bbb{R}\}$.
Hence we view the field $\phi$ rather as a path

$$\begin{array}{ccl}
\Bbb{R} & \longrightarrow & \mathfrak{F}\\
t & \longmapsto & [\vec{x}\longmapsto \phi(t,\vec{x}) = \Phi^{\vec{x}}(t)].
\end{array}$$
We thus recover the problem of studying the dynamics of a point moving in a configuration space $\mathfrak{F}$. The prices to
pay are 1) $\mathfrak{F}$ is infinite dimensional 2) we lose  relativistic invariance.\\

In this viewpoint, ${\cal L}[\phi] =
\int_{\Bbb{R}}\mathfrak{L}[t,\Phi(t),{d\Phi\over dt}(t)]dt$, where
$\Phi(t) = [\vec{x}\longmapsto \phi(t,\vec{x})]\in \mathfrak{F}$,
${d\Phi\over dt}(t) = [\vec{x}\longmapsto {\partial \phi\over \partial t}(t,\vec{x})]\in T_{\Phi(t)}\mathfrak{F}$ and
$\mathfrak{L}[t,\Phi(t),{d\Phi\over dt}(t)] = \int_{\Bbb{R}^{n-1}}L(x,\phi(x),d\phi(x))d\vec{x}$.\\

Then we consider the ``symplectic'' manifold which is formally $T^{\star}\mathfrak{F}$, i. e. we
introduce the dual variable

$$\Pi := {\partial \mathfrak{L}\over \partial {d\Phi\over dt}},$$
or equivalentely $\Pi(t) = [\vec{x}\longmapsto \pi(t,\vec{x})=\Pi_{\vec{x}}(t)]$ with

$$\Pi_{\vec{x}}(t) = {\partial \mathfrak{L}\over \partial {d\Phi^{\vec{x}}\over dt}}[t,\Phi(t),{d\Phi\over dt}(t)]\quad
\Longleftrightarrow \quad
\pi(t,\vec{x})={\delta \mathfrak{L}\over \delta {\partial \phi(t,\vec{x})\over \partial t}}[t,\Phi(t),{d\Phi\over dt}(t)]
= {\partial L\over \partial v_0}(x,\phi(x),d\phi(x)).$$
Here ${\delta \over \delta \phi(\vec{x})}$ is the Fr\'echet derivative. In our example

$$\Pi_{\vec{x}}(t) = {\partial \phi\over \partial t}(t,\vec{x}).$$

We define the Hamiltonian functional to be

$$\mathfrak{H}[\Phi,\Pi] := \int_{\Bbb{R}^{n-1}}\Pi_{\vec{x}}{\dot{\Phi}}^{\vec{x}}d\vec{x} - \mathfrak{L}[t,\Phi, {d\Phi\over dt}]
= \int_{\Bbb{R}^{n-1}}\left( {1\over 2}\pi(\vec{x})^2+{1\over 2}|\nabla \phi (\vec{x})|^2 + V(\phi(\vec{x}))\right)  d\vec{x}.$$

Now we can write the equations of motion as

$$\begin{array}{ccccl}
\displaystyle {\partial \pi\over \partial t}(t,\vec{x}) = {d\Pi_{\vec{x}}\over dt} & = &\displaystyle 
- {\partial \mathfrak{H}\over \partial \Phi^{\vec{x}}}(\Phi,\Pi)  & =
& \Delta \phi - V'(\phi)\\
\displaystyle {\partial \phi\over \partial t}(t,\vec{x}) = {d\Phi^{\vec{x}}\over dt} & = &
\displaystyle {\partial \mathfrak{H}\over \partial \Pi_{\vec{x}}}(\Phi,\Pi) & =
& \pi(t,\vec{x}).
\end{array}$$
A Poisson bracket can be defined on the set of functionals
$\{A:T^{\star}\mathfrak{F}\longmapsto \Bbb{R}\}$ by

$$\{A,B\} :=  \int_{\Bbb{R}^{n-1}}\left( {\delta A\over \delta \pi(\vec{x})}{\delta B\over \delta \phi(\vec{x})}
- {\delta A\over \delta \phi(\vec{x})}{\delta B\over \delta \pi(\vec{x})}\right) d\vec{x},$$
where ${\delta A\over \delta \phi(\vec{x})}$ is the Fr\'echet derivative with respect to $\phi(\vec{x})$, i. e. 
the distribution such that for any smooth compactly supported deformation $\delta \phi$ of $\phi$,

$$dA_{\phi}[\delta \phi] = \int_{\Bbb{R}^{n-1}}\delta \phi(\vec{x}){\delta A\over \delta \phi(\vec{x})}d\vec{x}.$$

And we may formulate the dynamical equations using the Poisson bracket as

$$\begin{array}{ccl}
\displaystyle {d\Pi_{\vec{x}}\over dt} & = & \displaystyle \{\mathfrak{H},\Pi_{\vec{x}}\}\\
\displaystyle {d\Phi^{\vec{x}}\over dt} & = &\displaystyle \{\mathfrak{H},\Phi^{\vec{x}}\},
\end{array}$$
with

$$\{\Phi^{\vec{x}},\Phi^{\vec{x}'}\} = \{\Pi_{\vec{x}},\Pi_{\vec{x}'}\} = 0,\quad
\{\Pi_{\vec{x}},\Phi^{\vec{x}'}\} = \delta_{\vec{x}}^{\vec{x}'} = \delta^{n-1}(\vec{x} - \vec{x}').$$
This singular Poisson bracket means that for any test functions $f,g\in {\cal C}^{\infty}_c(\Bbb{R}^{n-1},\Bbb{R})$,

$$\left\{ \int_{\Bbb{R}^{n-1}}g(\vec{x})\Pi_{\vec{x}}d\vec{x},
\int_{\Bbb{R}^{n-1}}f(\vec{x}')\Phi^{\vec{x}'}d\vec{x}'
\right\} = 
\int_{\Bbb{R}^{n-1}} f(\vec{x})g(\vec{x})d\vec{x}.$$
This implies in particular

$$\left\{ \int_{\Bbb{R}^{n-1}}g(\vec{x})\Pi_{\vec{x}}d\vec{x},
\int_{\Bbb{R}^{n-1}}f(\vec{x}')V(\Phi^{\vec{x}'})d\vec{x}'
\right\} = 
\int_{\Bbb{R}^{n-1}} V'(\Phi^{\vec{x}})f(\vec{x})g(\vec{x})d\vec{x},$$
because of the derivation property of the Poisson bracket.\\

\subsection{Translation in pataplectic geometry}

We first adapt and modify our notations: the coordinates on
${\cal M}=\Lambda^nT^{\star}(\Bbb{R}\times \Bbb{R}^{n-1}\times \Bbb{R})$
are now written $(q^{\mu},p_{\mu_1\dots \mu_n}) =  (x^{\alpha}, y,\epsilon,p^{\alpha} )$
where $0\leq \alpha \leq n-1$, $q^0=x^0=t$,
$(x^{\alpha})_{1\leq \alpha\leq n-1}=\vec{x}$, $q^n=y$ and

$$\epsilon:= p_{0\dots (n-1)}\quad p^{\alpha}:= p_{0\dots (\alpha-1)n(\alpha+1)\dots (n-1)}.$$
Hence

$$\theta = \epsilon\; dx^0\wedge \dots \wedge dx^{n-1} + \sum_{\alpha=0}^{n-1}
p^{\alpha}dx^0\wedge \dots \wedge dx^{\alpha-1}\wedge dy\wedge dx^{\alpha+1}\wedge \dots \wedge dx^{n-1},$$
or letting $\omega:= dx^0\wedge \dots \wedge dx^{n-1}$ and
$\omega_{\alpha}:= (-1)^{\alpha}dx^0\wedge \dots \wedge dx^{\alpha-1}\wedge dx^{\alpha+1}\wedge \dots \wedge dx^{n-1}
={\partial \over \partial x^{\alpha}}\inn \omega$,

$$\theta = \epsilon\; \omega + \sum_{\alpha=0}^{n-1}p^{\alpha}dy\wedge \omega_{\alpha}\quad
\hbox{and}\quad
\Omega = d\epsilon \wedge \omega + \sum_{\alpha=0}^{n-1}dp^{\alpha}\wedge dy\wedge \omega_{\alpha}.$$

Thus we see that in the present case the pataplectic formalism reduces essentially to the Weyl
formalism, because the fields are one dimensional.\\

Let us consider some field $\phi$ and a  map
$x\longmapsto p(x)$ such that $(x,\phi(x),d\phi(x))\leftrightarrow (x,\phi(x),p(x))$
\footnote{meaning that for some $w:\Bbb{R}\times \Bbb{R}^{n-1}\longrightarrow \Bbb{R}$,
we have $(x,\phi(x),d\phi(x),w(x))\leftrightarrow (x,\phi(x),p(x))$}. This implies the following relations

$$p^{\alpha} = {\partial L\over \partial v_{\alpha}}(x,\phi(x),d\phi(x))\quad \hbox{and} \quad
\epsilon = w +L(x,\phi(x),d\phi(x)) - \sum_{\alpha=0}^{n-1}p^{\alpha}{\partial \phi\over \partial x^{\alpha}}(x).$$

We let $\Gamma:= \{(x,\phi(x),p(x))/x\in \Bbb{R}\times \Bbb{R}^{n-1}\}
\subset {\cal M}$ and we consider the instantaneous
slices $S_t:= \Gamma \cap \{x^0=t\}$. These slices are oriented by the condition
${\partial \over \partial t}\inn \omega_{|S_t}>0$.
Then we can express the observables

$$\Phi^f(t):= \int_{\Bbb{R}^{n-1}}f(\vec{x})\Phi^{\vec{x}}(t)d\vec{x},\quad 
\Pi_g(t):= \int_{\Bbb{R}^{n-1}}g(\vec{x})\Pi_{\vec{x}}(t)d\vec{x}$$
and

$$\mathfrak{H}[\Phi(t),\Pi(t)] = \int_{\Bbb{R}^{n-1}}\left( \pi(t,\vec{x}){\partial \phi\over \partial t}(t,\vec{x})
-L(t,\vec{x},\phi(x),d\phi(x))\right) d\vec{x} =
\int_{\Bbb{R}^{n-1}}H^0_0(t,\vec{x},\phi)\omega_0$$
as integrals of $(n-1)$-forms on $S_t$. First

$$\Phi^f(t) = \int_{S_t}f(\vec{x})\phi(t,\vec{x})dx^1\wedge \dots \wedge dx^{n-1}
 = \int_{S_t}Q^f,\quad \hbox{with }Q^f:= f(\vec{x})\; y\; \omega_0.$$

$$\Pi_g(t) = \int_{S_t}g(\vec{x})\pi(t,\vec{x})dx^1\wedge \dots \wedge dx^{n-1} =
\int_{S_t}P_g,
\quad \hbox{with }P_g:= g(\vec{x})\sum_{\alpha=0}^{n-1}p^{\alpha}\omega_{\alpha},$$
because $\pi(t,\vec{x}) = {\partial L\over \partial v_0}(x,\phi(x),d\phi(x)) = p^0$
and $\omega_{\alpha|S_t} = 0$ if $\alpha\geq 1$\\

\noindent And last

$$\mathfrak{H}[\Phi(t),\Pi(t)] = \int_{\Bbb{R}^{n-1}}{\cal H}(q,p)\omega_0 -
\int_{\Bbb{R}^{n-1}}\epsilon \omega_0 +\sum_{\alpha=1}^{n-1}p^{\alpha}dy\wedge
\left( {\partial \over \partial x^{\alpha}}\inn \omega_0\right) =
\int_{S_t}\eta_0,$$
where

$$\eta_0:= {\cal H}(q,p)\omega_0 -\left( \epsilon \omega_0 +\sum_{\alpha=1}^{n-1}p^{\alpha}
dx^1\wedge \dots \wedge dx^{\alpha-1}\wedge dy\wedge dx^{\alpha+1}\wedge \dots \wedge dx^{n-1}\right) ,$$
because $H^0_0(x,\phi) = {\cal H}(q,p) -\langle p,{\partial \over \partial t}\wedge z_1\wedge \dots \wedge z_{n-1}\rangle
={\cal H}(q,p) - \left( \epsilon +\sum_{\alpha=1}^{n-1}p^{\alpha}{\partial \phi\over \partial x^{\alpha}}\right)$.

We remark \footnote{we observe also that
$P_g = g(\vec{x}){\partial \over \partial y}\inn \theta =
g(\vec{x}){\partial \over \partial y}\inn
(\theta - {\cal H}(q,p)\omega)$.} that

$$\eta_0 = - {\partial \over \partial t}\inn (\theta - {\cal H}(q,p)\omega).$$

\subsection{Recovering  the usual Poisson brackets as a local expression}

\noindent Our aim is now to express the various Poisson brackets involving
the quantities $\Phi^f(t)$ and $\Pi_g(t)$ along $\Gamma$ using some analogue of
the Poisson bracket defined on $(n-1)$-forms.
We generalize slightly the definition of $Q^f$ to be

\begin{equation}\label{3.3.f}
Q^f = \sum_{\alpha=0}^{n-1}f^{\alpha}(x)\; y\; \omega_{\alpha},
\end{equation}
where $f:= \sum_{\alpha=0}^{n-1}f^{\alpha}(x){\partial \over \partial x^{\alpha}}$
is some vector field. Hence our observables become

\begin{equation}\label{3.3.QP}
\Phi^f(t) = \int_{S_t}Q^f\quad \hbox{and}\quad \Pi_g(t) = \int_{S_t}P_g,
\end{equation}
where $P_g:= g(x)\sum_{\alpha=0}^{n-1}p^{\alpha}\omega_{\alpha}$ as before
\footnote{notice that
actually $\int_{S_t}Q^f = \int_{S_t}f^0(x)\; y\; \omega_0$.}.
We shall see here that we can define a bracket operation $\{.,.\}$ between
$Q^f$, $P_g$ and $\eta_0$ such that the usual Poisson bracket of fields
actually derives from $\{.,.\}$ by 

\begin{equation}\label{trans}
\int_{S_t}\{P_g,Q^f\} = \left\{ \int_{S_t}P_g,\int_{S_t}Q^f\right\},\;\hbox{etc}\dots
\end{equation}

\noindent First we remark that

$$\begin{array}{ccl}
dQ^f & = & \displaystyle \sum_{\alpha=0}^{n-1}f^{\alpha}\; dy\wedge \omega_{\alpha} +
\sum_{\alpha=0}^{n-1}y{\partial f^{\alpha}\over \partial x^{\alpha}}\omega 
= \sum_{\alpha=0}^{n-1}f^{\alpha}{\partial \over \partial p^{\alpha}}\inn \Omega +
\sum_{\alpha=0}^{n-1}y{\partial f^{\alpha}\over \partial x^{\alpha}}{\partial \over \partial \epsilon}\inn \Omega \\
& = & - \xi_{Q^f}\inn \Omega
\end{array}$$
and

$$\begin{array}{ccl}
dP_g & = & \displaystyle \sum_{\alpha=0}^{n-1}p^{\alpha}{\partial g\over \partial x^{\alpha}} \omega +
\sum_{\alpha=0}^{n-1}gdp^{\alpha}\wedge \omega_{\alpha} 
= \sum_{\alpha=0}^{n-1}p^{\alpha}{\partial g\over \partial x^{\alpha}} 
{\partial \over \partial \epsilon}\inn \Omega 
- g{\partial \over \partial y}\inn \Omega \\
& = & - \xi_{P_g}\inn \Omega,
\end{array}$$
where

\begin{equation}\label{3.3.A}
\xi_{Q^f}:= -\sum_{\alpha=0}^{n-1}f^{\alpha}{\partial \over \partial p^{\alpha}}
- y\sum_{\alpha=0}^{n-1}{\partial f^{\alpha}\over \partial x^{\alpha}}{\partial \over \partial \epsilon}
\end{equation}
and

\begin{equation}\label{3.3.B}
\xi_{P_g}:= g{\partial \over \partial y} -
\sum_{\alpha=0}^{n-1} p^{\alpha}{\partial g\over \partial x^{\alpha}}{\partial \over \partial \epsilon}.
\end{equation}
Also notice that

$$d\eta_0 = (d{\cal H}-d\epsilon)\wedge \omega_0 - \sum_{\alpha=1}^{n-1}dp^{\alpha}\wedge dy\wedge
\left( {\partial \over \partial x^{\alpha}}\inn \omega_0\right) .$$

\begin{defi} We define the Poisson $\mathfrak{p}$-brackets of these $(n-1)$-forms to be

$$\{\eta_0,Q^f\} := - \xi_{Q^f}\inn d\eta_0,\quad \{\eta_0,P_g\} := - \xi_{P_g}\inn d\eta_0,$$

$$\{P_g,Q^f\} := - \xi_{Q^f}\inn dP_g =  \xi_{P_g}\inn dQ^f  = \xi_{Q^f}\inn (\xi_{P_g}\inn \Omega)$$
and
$$\{Q^f,Q^{f'}\} := \xi_{Q^{f'}}\inn (\xi_{Q^f}\inn \Omega),\quad 
\{P_g,P_{g'}\} := \xi_{P_{g'}}\inn (\xi_{P_g}\inn \Omega).$$
\end{defi}

\noindent Let us now compute these $\mathfrak{p}$-brackets. We use in particular
the fact that ${\partial {\cal H}\over \partial \epsilon} = 1$.

$$\begin{array}{ccl}
\{\eta_0,Q^f\} & = & \displaystyle \left( \sum_{\alpha=0}^{n-1}f^{\alpha}{\partial \over \partial p^{\alpha}}
+ y\sum_{\alpha=0}^{n-1}{\partial f^{\alpha}\over \partial x^{\alpha}}{\partial \over \partial \epsilon}\right)
\inn \left( (d{\cal H}-d\epsilon)\wedge \omega_0 - \sum_{\alpha=1}^{n-1}dp^{\alpha}\wedge dy\wedge \left( {\partial \over \partial x^{\alpha}}\inn \omega_0\right) \right) \\
& = & \displaystyle \sum_{\alpha=0}^{n-1}f^{\alpha}{\partial {\cal H}\over \partial p^{\alpha}}\omega_0
- \sum_{\alpha=1}^{n-1}f^{\alpha}dy\wedge 
\left( {\partial \over \partial x^{\alpha}}\inn \omega_0\right) .
\end{array}$$

$$\begin{array}{ccl}
\{\eta_0,P_g\} & = & \displaystyle \left(
\sum_{\alpha=0}^{n-1} p^{\alpha}{\partial g\over \partial x^{\alpha}}{\partial \over \partial \epsilon}
- g{\partial \over \partial y}\right) 
\inn \left( (d{\cal H}-d\epsilon)\wedge \omega_0 - \sum_{\alpha=1}^{n-1}dp^{\alpha}\wedge dy\wedge \left( {\partial \over \partial x^{\alpha}}\inn \omega_0\right) \right) \\
& = & \displaystyle -g{\partial {\cal H}\over \partial y}\omega_0 - g\sum_{\alpha=1}^{n-1}dp^{\alpha}\wedge \left( {\partial \over \partial x^{\alpha}}\inn \omega_0\right) ,
\end{array}$$

$$\begin{array}{ccl}
\{P_g,Q^f\} & = & \displaystyle \left(
\sum_{\alpha=0}^{n-1}f^{\alpha}{\partial \over \partial p_{\alpha}} +
y\sum_{\alpha=0}^{n-1}{\partial f^{\alpha}\over \partial x^{\alpha}}{\partial \over \partial \epsilon}
\right) 
\inn \left( \sum_{\alpha=0}^{n-1} p^{\alpha}{\partial g\over \partial x^{\alpha}}\omega
- g{\partial \over \partial y}\inn \Omega\right) \\
& = & \displaystyle  g\sum_{\alpha=0}^{n-1}f^{\alpha}\omega_{\alpha},
\end{array}$$
and $\{Q^f,Q^{f'}\} = \{P_g,P_{g'}\} = 0$.
We now integrate the $\mathfrak{p}$-brackets on a constant time slice $S_t\subset \Gamma$. We 
immediately see that

$$\int_{S_t}\{P_g,Q^f\} =   \int_{S_t}g\; f^0\omega_0
= \{\pi_g(t),\Phi^f(t)\} = \left\{ \int_{S_t}P_g,\int_{S_t}Q^f\right\}$$
and we recover (\ref{trans}). Second,

$$\int_{S_t}\{\eta_0,Q^f\} = \int_{S_t}\sum_{\alpha=0}^{n-1}f^{\alpha}
{\partial {\cal H}\over \partial p^{\alpha}}\omega_0
- \sum_{\alpha=1}^{n-1}f^{\alpha}{\partial \phi\over \partial x^{\alpha}} \omega_0.$$
Third,

$$\int_{S_t}\{\eta_0,P_g\} =  \int_{S_t}-g{\partial {\cal H}\over \partial y}\omega_0
- \sum_{\alpha=1}^{n-1}g{\partial p^{\alpha}\over \partial x^{\alpha}}\omega_0.
$$
Now let us assume that $\Gamma$ is the graph of a solution of the Hamilton equations
(\ref{hamilton3}) or (\ref{weyl}).
Since then ${\partial \phi\over \partial x^{\alpha}}
= {\partial {\cal H}\over \partial p^{\alpha}}$ along $\Gamma$,

$$\int_{S_t}\{\eta_0,Q^f\} = \int_{S_t}f^0{\partial \phi\over \partial t}\omega_0,$$
and because of $-{\partial {\cal H}\over \partial y}
- \sum_{\alpha=1}^{n-1}{\partial p^{\alpha}\over \partial x^{\alpha}} =
{\partial p^0\over \partial t}$,

$$\int_{S_t}\{\eta_0,P_g\} = \int_{S_t}g{\partial p^0\over \partial t}\omega_0.$$
We conclude that

$${d\over dt}\int_{S_t}Q^f =
{d\over dt}\Phi^f(t) = \int_{S_t}f^0{\partial \phi\over \partial t}\omega_0
+{\partial f^0\over \partial t}\phi\omega_0
= \int_{S_t}\{\eta_0,Q^f\} + \Phi^{\partial f/\partial t}(t) $$
and

$${d\over dt}\int_{S_t}P_g =
{d\over dt}\Pi_g(t) = \int_{S_t}g{\partial p^0\over \partial t}\omega_0 +
{\partial g\over \partial t}p^0\omega_0
= \int_{S_t}\{\eta_0,P_g\} + \Pi_{\partial g/\partial t}(t).$$
This has to be compared with the usual canonical equations for fields:

$${d\over dt}\int_{S_t}Q^f =
\left\{\int_{S_t}\eta_0,\int_{S_t}Q^f\right\} + \Phi^{\partial f/\partial t}(t)
\quad \hbox{and}\quad
{d\over dt}\int_{S_t}P_g =
\left\{\int_{S_t}\eta_0,\int_{S_t}P_g\right\} + \Pi_{\partial g/\partial t}(t).$$

\subsection{An alternative dynamical formulation using $\mathfrak{p}$-brackets}

\noindent We can also define the $\mathfrak{p}$-bracket of a $n$-form with forms
$Q^f$ or $P_g$ as given by (\ref{3.3.f}) and (\ref{3.3.QP}).
If $\psi$ is such a $n$-form, 

$$\{ \psi, Q^f\} := -\xi_{Q^f}\inn d\psi\quad
\hbox{and}\quad \{ \psi, P_g\} := -\xi_{P_g}\inn d\psi,$$
where (\ref{3.3.A}) and (\ref{3.3.B}) have been used.
An important instance is for $\psi = {\cal H}\omega$:

$$\{ {\cal H}\omega, Q^f\} = \sum_{\alpha=0}^{n-1}f^{\alpha}{\partial {\cal H}\over \partial p^{\alpha}}\omega
+ \sum_{\alpha=0}^{n-1}y{\partial f^{\alpha}\over \partial x^{\alpha}}\omega.$$
We shall integrate this $\mathfrak{p}$-bracket on
$\Gamma^{t_2}_{t_1}:= \{(q,p)\in \Gamma/t_1<t<t_2\}$, where we still assume that
$\Gamma$ is the graph of a solution of the Hamilton equations (\ref{hamilton6}).
An integration by parts gives 

$$\begin{array}{ccl}
\displaystyle \int_{\Gamma_{t_1}^{t_2}}\{ {\cal H}\omega, Q^f\} & = &\displaystyle 
\int_{\partial \Gamma_{t_1}^{t_2}}\phi \sum_{\alpha=0}^{n-1}f^{\alpha}\omega_{\alpha}
+ \int_{\Gamma_{t_1}^{t_2}}\sum_{\alpha=0}^{n-1}f^{\alpha}\left( {\partial {\cal H}\over \partial p^{\alpha}}
- {\partial \phi\over \partial x^{\alpha}}\right) \omega\\
& = & \displaystyle \int_{\partial \Gamma_{t_1}^{t_2}}Q^f = \int_{S_{t_2}}Q^f - \int_{S_{t_1}}Q^f.
\end{array}$$
Similarly we find that

$$\{ {\cal H}\omega, P_g\} = \sum_{\alpha=0}^{n-1}p^{\alpha}{\partial g\over \partial x^{\alpha}}\omega
- g{\partial {\cal H}\over \partial y}\omega,$$
and thus

$$\begin{array}{ccl}
\displaystyle \int_{\Gamma_{t_1}^{t_2}}\{ {\cal H}\omega, P_g\} & = &\displaystyle 
\int_{\partial \Gamma_{t_1}^{t_2}}\sum_{\alpha=0}^{n-1}gp^{\alpha}\omega_{\alpha}
- \int_{\Gamma_{t_1}^{t_2}}g\left( {\partial {\cal H}\over \partial y}
+ \sum_{\alpha=0}^{n-1}{\partial p^{\alpha}\over \partial x^{\alpha}}\right) \omega \\
& =& \displaystyle \int_{\partial \Gamma_{t_1}^{t_2}}P_g = \int_{S_{t_2}}P_g - \int_{S_{t_1}}P_g.
\end{array}$$
We are tempted to conclude that

$${\bf d}Q^f = \{ {\cal H}\omega, Q^f\}\quad \hbox{and}\quad 
{\bf d}P_g = \{ {\cal H}\omega, P\},$$
where ${\bf d}$ is the differential along a graph $\Gamma$ of a solution of the Hamilton equations
(\ref{hamilton3}). This precisely will be proven in the next section.

\section{Poisson $\mathfrak{p}$-brackets for $(p-1)$-forms on ${\cal M}$}
We have seen on some examples that the Poisson bracket algebra of the classical field theory
can actually be derived from brackets on $(n-1)$-forms which are integrated
on constant time slices. Actually these constructions can be generalized in
several ways.

\subsection{$\mathfrak{p}$-brackets on $(n-1)$-forms}
We turn back to ${\cal M} = \Lambda^nT^{\star}({\cal X}\times {\cal Y})$ and
to the notation of the previous Section.
Let $\Gamma({\cal M},\Lambda^{n-1}T^{\star}{\cal M})$ be the set of smooth $(n-1)$-forms
on ${\cal M}$.  We consider the subset $\mathfrak{P}^{n-1}{\cal M}$ of
$\Gamma({\cal M},\Lambda^{n-1}T^{\star}{\cal M})$ of
forms $a$ such that there exists a vector field $\xi_{a}=\Xi(a)$
which satisfies the property

$$da = - \xi_{a}\inn \Omega.$$
Obviously $\Xi({a})$ depends only on $a$ modulo closed forms and the
map $a\longmapsto \Xi({a})$ from $\mathfrak{P}^{n-1}{\cal M}$
to the set of vector fields induces a map on the quotient
$\mathfrak{P}^{n-1}{\cal M}/C^{n-1}({\cal M})$, where $C^{n-1}({\cal M})$
is the set of closed $(n-1)$-forms.
A property of vector fields $\Xi(a)$ is that there are infinitesimal
symmetries of $\Omega$, for

$${\cal L}_{\Xi(a)}\Omega = d\left( \Xi(a)\inn \Omega\right)
+ \Xi(a)\inn d\Omega = -d\circ da = 0.$$
We shall denote $\mathfrak{pp}{\cal M}$ the set of pataplectic vector fields, i.e.
vector fields $X$ such that $X\inn \Omega$ is exact. Clearly
$\Xi:\mathfrak{P}^{n-1}{\cal M}/C^{n-1}({\cal M})\longrightarrow \mathfrak{pp}{\cal M}$
is a vector space isomorphism.

Then we define the {\em internal $\mathfrak{p}$-bracket} on $\mathfrak{P}^{n-1}{\cal M}$ by

$$\{a,b\} := \Xi({b})\inn \Xi(a)\inn \Omega.$$
\begin{lemm}
For any $a,b\in \mathfrak{P}^{n-1}{\cal M}$,
\begin{equation}\label{poissonlie}
d\{a,b\} = - [\Xi({a}),\Xi({b})]\inn \Omega.
\end{equation}
\end{lemm}
{\bf Proof} Let $\xi_{a}=\Xi(a)$ and $\xi_{b}=\Xi(b)$. Then denoting
${\cal L}_{\xi_{a}}$ the Lie derivative with respect to $\xi_{a}$,

$$\begin{array}{ccl}
[\xi_{a},\xi_{b}] \inn \Omega & = & {\cal L}_{\xi_{a}}(\xi_{b})\inn \Omega\\
 & = &{\cal L}_{\xi_{a}}\left( \xi_{b}\inn \Omega\right) -
\xi_{b}\inn {\cal L}_{\xi_{a}}(\Omega)\\
 & = & d(\xi_{a}\inn \xi_{b}\inn \Omega) + \xi_{a}\inn d(\xi_{b}\inn \Omega)
- \xi_{b}\inn (d(\xi_{a}\inn \Omega) + \xi_{a}\inn d\Omega).
\end{array}$$
But since $d\Omega = d(\xi_{a}\inn \Omega) = d(\xi_{b}\inn \Omega) = 0$, we find that
$[\xi_{a},\xi_{b}]\inn \Omega = d(\xi_{a}\inn \xi_{b}\inn \Omega)
= -d\{a,b\}$. \bbox

\begin{lemm}
$\Xi:\mathfrak{P}^{n-1}{\cal M}/C^{n-1}({\cal M}) \longrightarrow \mathfrak{pp}{\cal M}$
is a Lie algebra isomorphism. More precisely we have

\begin{equation}\label{isomorphism}
\Xi(\{a,b\}) = [\Xi({a}),\Xi({b})].
\end{equation}
This implies the Jacobi identity modulo exact terms in $\mathfrak{P}^{n-1}{\cal M}$:

\begin{equation}\label{jacobi}
\{\{a,b\},c\} + \{\{b,c\},a\} + \{\{c,a\},b\} = d(\xi_c\inn \xi_b\inn \xi_a\inn \Omega).
\end{equation}
\end{lemm}
{\bf Proof} Relation (\ref{isomorphism}) is a direct consequence of (\ref{poissonlie}) in
Lemma 2. The Jacobi identity follows from

$$\begin{array}{ccl}
\{\{a,b\},c\} & = & \xi_c\inn [\xi_a,\xi_b]\inn \Omega\\
 & = & \xi_c\inn d(\xi_a\inn \xi_b\inn \Omega)\\
 & = & {\cal L}_{\xi_c}(\xi_a\inn \xi_b\inn \Omega) - d(\xi_c\inn \xi_a\inn \xi_b\inn \Omega)\\
 & = & [\xi_c,\xi_a]\inn \xi_b\inn \Omega + \xi_a\inn [\xi_c,\xi_b]\inn \Omega
+ \xi_a\inn \xi_b\inn {\cal L}_{\xi_c}(\Omega) + d(\xi_c\inn \xi_b\inn \xi_a\inn \Omega)\\
 & = & - \{\{c,a\},b\} - \{\{b,c\},a\} + d(\xi_c\inn \xi_b\inn \xi_a\inn \Omega),
\end{array}$$
where we have used (\ref{isomorphism}). \bbox

We can extend the definition of the $\mathfrak{p}$-bracket: for any
$0\leq p\leq n$ the {\em external $\mathfrak{p}$-bracket} of a $p$-form
$a\in \Gamma({\cal M},\Lambda^pT^{\star}{\cal M})$
with a form $b\in \mathfrak{P}^{n-1}{\cal M}$ is

$$\{a,b\} = - \{b,a\}:= - \Xi(b)\inn da.$$
Of course this definition coincides with the previous one when
$a\in \mathfrak{P}^{n-1}{\cal M}$.\\

\noindent {\bf Examples of external $\mathfrak{p}$-brackets}
For any $a\in \mathfrak{P}^{n-1}{\cal M}$,

$$\{\theta,a\} = -\Xi(a)\inn d\theta = - \Xi(a)\inn \Omega
= da.$$
We can add that it is worthwhile to write in the external $\mathfrak{p}$-brackets
of observable forms
like $q^{\mu}$, $q^{\mu}dq^{\nu}$, etc ...
$$\begin{array}{ccl}
\{P_{i,g},q^{\mu}\} & = & \Xi(P_{i,g}) \inn dq^{\mu} = g \delta^{\mu}_i,\\
\{Q^{i,f},q^{\mu}\} & = & \Xi(Q^{i,f}) \inn dq^{\mu} = 0\\
\{P_{i,g},q^{\mu}dq^{\nu}\} & = & \Xi(P_{i,g}) \inn dq^{\mu}\wedge dq^{\nu} =
g\left(  \delta^{\mu}_idq^{\nu} - \delta^{\nu}_idq^{\mu}\right) .
\end{array}$$

\begin{theo}
Let $\Gamma$ be the graph in ${\cal M}$ of a solution of the Hamilton equations
(\ref{hamilton3}) and write ${\cal U}:x\longmapsto {\cal U}(x) = (x,u(x),p(x))$ the natural
parametrization of $\Gamma$. Then for any form $a\in \mathfrak{P}^{n-1}{\cal M}$,

$${\bf d}a = \{{\cal H}\omega,a\},$$
where ${\bf d}$ is the differential along $\Gamma$
(meaning that $da_{|\Gamma} = \{{\cal H}\omega,a\}_{|\Gamma}$).
\end{theo}
{\bf Proof} We choose an arbitrary open subset $D\subset \Gamma$ and 
denoting $\xi_{a} = \Xi(a)$, we compute

$$\begin{array}{ccl}
\displaystyle \int_D\{{\cal H}\omega,a\} & = & \displaystyle 
- \int_D\xi_{a}\inn (d{\cal H}\wedge \omega)\\
& = & \displaystyle - \int_{{\cal U}^{-1}(D)}d{\cal H}\wedge \omega
\left( \xi_{a},{\partial {\cal U}\over \partial x^1},\dots ,
{\partial {\cal U}\over \partial x^n}\right) 
\omega\\
& = & \displaystyle - \int_{{\cal U}^{-1}(D)}(-1)^n
{\partial {\cal U}\over \partial x^1\dots \partial x^n}\inn (d{\cal H}\wedge \omega)(\xi_a)\omega.
\end{array}$$

We use  equation (\ref{hamilton6}) and obtain

$$\begin{array}{ccl}
\displaystyle \int_D\{{\cal H}\omega,a\} & = & \displaystyle 
- \int_{{\cal U}^{-1}(D)}(-1)^n{\partial {\cal U}\over \partial x^1\dots \partial x^n}\inn
\Omega(\xi_{a})\omega\\
& = & \displaystyle 
- \int_{{\cal U}^{-1}(D)}\Omega\left( \xi_{a},
{\partial {\cal U}\over \partial x^1},\dots ,{\partial {\cal U}\over \partial x^n}\right) 
\omega\\
& = & \displaystyle 
- \int_D\xi_{a}\inn \Omega = \int_Dda.
\end{array}$$
And the Theorem follows. \bbox
Another way to state this result is that

\begin{equation}\label{stokes}
\int_D\{{\cal H}\omega,a\} = \int_{\partial D}a
\end{equation}
along any solution of (\ref{hamilton3}).

\subsection{Expression of the standard observable $(n-1)$-forms}

\noindent These quantities are integrals of $(n-1)$-forms on hypersurfaces
which are thought as ``constant time slices'', the transversal dimension being
then considered as a local time. The target coordinates observables
\footnote{\label{position}comparing with the one-dimensional Hamiltonian formalism we can
see these target coordinates as generalizations of the position observables.}
are weighted integrals
of the value of the field and are induced by the {\em ``position'' $\mathfrak{p}$-forms}

$$Q^{i,f} := y^i\; \sum_{\alpha}f^{\alpha}(x) \omega_{\alpha} = y^i\; f\inn \omega,$$
where $f=\sum_{\alpha}f^{\alpha}(x){\partial \over \partial x^{\alpha}}$ is a
tangent vector field on ${\cal X}$ and
$\omega_{\alpha} = {\partial \over \partial x^{\alpha}}\inn \omega$. The ``momentum''
and ``energy'' observables are obtained from the {\em momentum form}

$$P_{\mu,g}^{\star}:=
g(x) {\partial \over \partial q^{\mu}}\inn (\theta-{\cal H}(q,p)\omega),$$
where $g$ is a smooth function on ${\cal X}$. Alternatively we may sometimes prefer to use
the $\mathfrak{p}$-forms

$$P_{\mu,g}:= g(x) {\partial \over \partial q^{\mu}}\inn \theta.$$
For $1\leq \mu=\alpha\leq n$,
$P_{\mu,g}^{\star} =: H_{\alpha,g}$ generates the components of the Hamiltonian
tensor but $P_{\alpha,g}$ (which is different from $P_{\alpha,g}^{\star}$)
does not in general.
However the restrictions of $P_{\mu,g}^{\star}$
and $P_{\mu,g}$ on the hypersurface ${\cal H}=0$ coincide so that if we work
on this hypersurface both forms can be used.
For $n+1\leq \mu=n+i\leq n+k$, $P_{\mu,g}^{\star} =P_{\mu,g} =: P_{i,g}$ generates the
momentum components
\footnote{ The advantage
of $P_{\mu,g}$ with respect to $P_{\mu,g}^{\star}$ is that
$P_{\mu,g}$ belongs to $\mathfrak{P}^{n-1}{\cal M}$ for all values of $\mu$.}.\\

To check that, we consider a parametrization
${\cal U}:x\longmapsto (x,u(x),p(x))$ of some graph $\Gamma$ and look at
the pull-back of these forms by ${\cal U}$. We write
${\cal U}^{\star}P_{\mu,g}^{\star} = \sum_{\beta}s^{\beta}\omega_{\beta}$, which implies
$s^{\beta}\omega= dx^{\beta}\wedge {\cal U}^{\star}P_{\mu,g}^{\star}$ and we compute
$$\begin{array}{ccl}
s^{\beta} & = & \displaystyle g(x)\left\langle p,{\partial q\over \partial x^1}\wedge \dots
\wedge {\partial q\over \partial x^{\beta-1}}\wedge
{\partial \over \partial q^{\mu}}\wedge {\partial q\over \partial x^{\beta+1}}
\wedge \dots \wedge {\partial q\over \partial x^n}\right\rangle
_{|z={\partial U\over \partial x}}
- g(x)\delta^{\beta}_{\mu}{\cal H}\\
& = &\displaystyle g(x){\partial \langle p,z\rangle \over \partial z^{\mu}_{\beta}}
_{|z={\partial U\over \partial x}}
- g(x)\delta^{\beta}_{\mu}{\cal H}.
\end{array}$$

Hence we find that

$${\cal U}^{\star}H_{\alpha,g} = -g(x)\sum_{\beta}H^{\beta}_{\alpha}(q(x),p(x))
\omega_{\beta} = g(x)\sum_{\beta}S^{\beta}_{\alpha}(x,u(x),du(x))\omega_{\beta},$$
$${\cal U}^{\star}P_{i,g} = g(x)\sum_{\beta}
{\partial \langle p,z\rangle \over \partial z^i_{\beta}}_{|z={\partial U\over \partial x}}
\omega_{\beta} = g(x)\sum_{\beta}{\partial L\over \partial v^i_{\beta}}(x,u(x),du(x))
\omega_{\beta}.$$
We shall prove below that $P_{\mu,g}$ (and hence $P_{i,g}$) and $Q^{i,f}$ belong to
$\mathfrak{P}^{n-1}{\cal M}$.\\

\subsubsection{Larger classes of observable}

These forms, which are enough to translate most of the observable
studied in the usual field theory, are embedded in two more general classes of
observables the definition of which follows.\\

\noindent {\bf Generalised positions} (see the footnote \ref{position})
They are forms $Q^{\zeta}$ in $\Lambda^{n-1}T^{\star}({\cal X}\times {\cal Y})$, i. e. 

$$Q^{\zeta}:= 
\sum_{\mu_1<\dots <\mu_{n-1}}
\zeta_{\mu_1\dots \mu_{n-1}}(q)
dq^{\mu_1}\wedge \dots \wedge dq^{\mu_{n-1}}. $$
An example is $Q^{i,f} = y^if(x)\partial_{\alpha}\inn \omega$.
We denote $\mathfrak{P}_Q^{n-1}{\cal M} =
\Lambda^{n-1}T^{\star}({\cal X}\times {\cal Y})$.\\

\noindent {\bf Generalised momenta} For each section of
$T({\cal X}\times{\cal Y})$, i. e. a vector field

$$\xi:= \sum_{\mu}\xi^{\mu}(q){\partial \over \partial q^{\mu}},$$
we define the $(n-1)$-form

$$P_{\xi}:= \xi\inn \theta =
\sum_{\mu}\xi^{\mu}{\partial \over \partial q^{\mu}}\inn \theta.$$
An example is for $\xi = g(x){\partial \over \partial q^{\mu}}$, then we
obtain $P_{g(x){\partial \over \partial q^{\mu}}} = P_{\mu,g}$.
We denote $\mathfrak{P}_P^{n-1}{\cal M}$ the set of such $(n-1)$-forms.

\begin{lemm}
$\mathfrak{P}^{n-1}_Q{\cal M}$ ans $\mathfrak{P}^{n-1}_P{\cal M}$ are subsets of
$\mathfrak{P}^{n-1}{\cal M}$, precisely
$$\begin{array}{ccl}
\Xi(Q^{\zeta}) & = & \displaystyle  \sum_{\mu_1<\dots <\mu_n}\sum_{\alpha}(-1)^{\alpha}
{\partial \zeta_{\mu_1\dots \mu_{\alpha-1}\mu_{\alpha+1}\dots \mu_n}
\over \partial q^{\mu_{\alpha}}}
{\partial \over \partial p_{\mu_1\dots \mu_n}},\\
 & & \\
\Xi(P_{\xi}) & = & \displaystyle \xi - \sum_{\mu}\sum_{\nu}
{\partial \xi^{\mu}\over \partial q^{\nu}}\Pi^{\nu}_{\mu},
\end{array}$$
where

$$\Pi^{\nu}_{\mu}:= \sum_{\mu_1<\dots <\mu_n}\sum_{\alpha}
p_{\mu_1\dots \mu_{\alpha-1}\mu\mu_{\alpha+1}\dots \mu_n}
\delta^{\nu}_{\mu_{\alpha}}{\partial \over \partial p_{\mu_1\dots \mu_n}}$$
so that 

$$dq^{\nu}\wedge {\partial \over \partial q^{\mu}}\inn \theta = 
\Pi^{\nu}_{\mu}\inn \Omega.$$
\end{lemm}
{\bf Proof} We have

$$\begin{array}{ccl}
dQ^{\zeta} & = & \displaystyle \sum_{\nu}\sum_{\mu_1<\dots <\mu_{n-1}}
{\partial \zeta_{\mu_1\dots  \mu_{n-1}}\over \partial q^{\nu}}
dq^{\nu}\wedge dq^{\mu_1}\wedge \dots \wedge dq^{\mu_{n-1}}\\
 & = & \displaystyle \sum_{\alpha}\sum_{\mu_1<\dots <\mu_n}
{\partial \zeta_{\mu_1\dots \mu_{\alpha-1}\mu_{\alpha+1}\dots \mu_n}
\over \partial q^{\mu_{\alpha}}}
dq^{\mu_{\alpha}}\wedge dq^{\mu_1}\wedge \dots \wedge
dq^{\mu_{\alpha-1}}\wedge dq^{\mu_{\alpha+1}}\wedge \dots \wedge dq^{\mu_n}\\
 & = & \displaystyle \sum_{\alpha}(-1)^{\alpha-1}\sum_{\mu_1<\dots <\mu_n}
{\partial \zeta_{\mu_1\dots \mu_{\alpha-1}\mu_{\alpha+1}\dots \mu_n}
\over \partial q^{\mu_{\alpha}}}
{\partial \over \partial p_{\mu_1\dots \mu_n}}\inn \Omega.
\end{array}$$
And the expression for $\Xi(Q^{\zeta})$ follows.\\

\noindent Next we write

$$dP_{\xi} = \sum_{\mu}\sum_{\nu}{\partial \xi^{\mu}\over \partial q^{\nu}}
dq^{\nu}\wedge {\partial \over \partial q^{\mu}}\inn \theta 
- \sum_{\mu}\xi^{\mu}{\partial \over \partial q^{\mu}}\inn \Omega$$
and we conclude by computing
$dq^{\nu}\wedge {\partial \over \partial q^{\mu}}\inn \theta$, indeed

$$\begin{array}{ccl}
\displaystyle dq^{\nu}\wedge {\partial \over \partial q^{\mu}}\inn \theta
& = & \displaystyle \sum_{\mu_1<\dots <\mu_n}\sum_{\alpha}
p_{\mu_1\dots \mu_n}\delta^{\mu_{\alpha}}_{\mu}
dq^{\mu_1}\wedge \dots \wedge dq^{\mu_{\alpha-1}}\wedge dq^{\nu}\wedge
dq^{\mu_{\alpha+1}}\wedge \dots \wedge dq^{\mu_n}\\
 & = &  \displaystyle \sum_{\mu_1<\dots <\mu_n}\sum_{\alpha}
p_{\mu_1\dots \mu_{\alpha-1}\mu\mu_{\alpha+1}\dots \mu_n}\delta^{\nu}_{\mu_{\alpha}}
dq^{\mu_1}\wedge \dots \wedge dq^{\mu_n}\\
 & = &  \displaystyle \sum_{\mu_1<\dots <\mu_n}\sum_{\alpha}
p_{\mu_1\dots \mu_{\alpha-1}\mu\mu_{\alpha+1}\dots \mu_n}\delta^{\nu}_{\mu_{\alpha}}
{\partial \over \partial p_{\mu_1\dots \mu_n}}\inn \Omega.
\end{array}$$
Hence we deduce the result on $P_{\xi}$.\bbox 

\noindent {\bf Poisson $\mathfrak{p}$-brackets}\\

\noindent We are now in position to compute the $\mathfrak{p}$-brackets of these
forms. The results are summarized in the following Proposition.

\begin{prop}
The $\mathfrak{p}$-brackets of forms in $\mathfrak{P}_Q^{n-1}{\cal M}$ and
$\mathfrak{P}_P^{n-1}{\cal M}$
are the following

$$\begin{array}{ccl}
\{Q^{\zeta}, Q^{\tilde{\zeta}}\} & = & 0\\
 & & \\
\{P_{\xi}, P_{\tilde{\xi}}\} & = & 
P_{[\xi,\tilde{\xi}]}+ d(\tilde{\xi}\inn \xi \inn \theta)\\
 & & \\
\{P_{\xi}, Q^{\zeta}\} & = &\displaystyle 
\sum_{\mu_1<\dots <\mu_n}\sum_{\alpha}\sum_{\mu}(-1)^{\alpha+1}\xi^{\mu}
{\partial \zeta_{\mu_1\dots \mu_{\alpha-1}\mu_{\alpha-1}\dots \mu_n}
\over \partial q^{\mu_{\alpha}}}
{\partial \over \partial q^{\mu}}\inn
dq^{\mu_1}\wedge \dots \wedge dq^{\mu_n}.
\end{array}$$
\end{prop}
{\bf Proof} These results are all straighforward excepted for $\{P_{\xi}, P_{\tilde{\xi}}\}$,.
We remark that $\Xi(P_{\xi})\inn \theta = \xi\inn \theta = P_{\xi}$ and

\begin{equation}\label{4.2.1.lie}
\begin{array}{ccl}
{\cal L}_{\Xi(P_{\xi})}(\theta) & = & \Xi(P_{\xi})\inn d\theta + d(\Xi(P_{\xi})\inn \theta)\\
 & = & \Xi(P_{\xi})\inn \Omega + dP_{\xi} = 0,
\end{array}
\end{equation}
so that $\Xi(P_{\xi})$ may be viewed as the extension of $\xi$ to a vector
field leaving $\theta$ invariant. Now we deduce that

$$\begin{array}{ccl}
[\xi, \tilde{\xi}]\inn \theta & = & [\Xi(P_{\xi}), \Xi(P_{\tilde{\xi}})]\inn \theta\\
 & = & {\cal L}_{\Xi(P_{\xi})}(\Xi(P_{\tilde{\xi}}))\inn \theta\\
 & = & {\cal L}_{\Xi(P_{\xi})}(\Xi(P_{\tilde{\xi}})\inn \theta)
- \Xi(P_{\tilde{\xi}})\inn {\cal L}_{\Xi(P_{\xi})}\theta\\
 & = & \Xi(P_{\xi})\inn d(\Xi(P_{\tilde{\xi}})\inn \theta) +
d (\Xi(P_{\xi})\inn \Xi(P_{\tilde{\xi}})\inn \theta)
- \Xi(P_{\tilde{\xi}})\inn 0\\
 & = & \Xi(P_{\xi})\inn dP_{\tilde{\xi}} + d(\xi\inn \tilde{\xi}\inn \theta)\\
 & = & - \Xi(P_{\xi})\inn \Xi(P_{\tilde{\xi}})\inn \Omega + d(\xi\inn \tilde{\xi}\inn \theta)\\
 & = & \{P_{\xi},P_{\tilde{\xi}}\} - d(\tilde{\xi}\inn \xi\inn \theta)
\end{array}$$
And the result follows. \bbox

\subsubsection{Back to the standard observables}

As an application of the previous results we can express the pataplectic vector
fields associated to $Q^{i,f}$ and $P_{\mu,g}$ and their $\mathfrak{p}$-brackets.
For that purpose, it is useful to introduce other notations:

$$\begin{array}{ccl}
\epsilon & := & p_{1\dots n}\\
p^{\alpha}_i & := & p_{1\dots (\alpha-1)(n+i)(\alpha+1)\dots n}\\
p^{\alpha_1\alpha_2}_{i_1i_2} & := &
p_{1\dots (\alpha_1-1)(n+i_1)(\alpha_1+1)\dots 
(\alpha_2-1)(n+i_2)(\alpha_2+1)\dots n}\\
& & \hbox{etc}\dots 
\end{array}$$
and
$$\begin{array}{ccl}
\omega^i_{\alpha} & := & dy^i\wedge \left( {\partial \over \partial x^{\alpha}}
\inn \omega\right) =: (dy^i\wedge \partial _{\alpha})\inn \omega\\
\omega^{i_1i_2}_{\alpha_1\alpha_2} & := & (dy^{i_1}\wedge \partial _{\alpha_1})\inn 
(dy^{i_2}\wedge \partial _{\alpha_2})\inn  \omega\\
& & \hbox{etc}\dots 
\end{array}$$
in such a way that

$$\theta = \epsilon\;\omega  + \sum_{p=1}^n{1\over p!^2}
\sum_{i_1,\dots ,i_p;\alpha_1,\dots ,\alpha_p}
p^{\alpha_1\dots \alpha_p}_{i_1\dots i_p}
\omega^{i_1\dots i_p}_{\alpha_1\dots \alpha_p}.$$
(Notice that the Weyl theory corresponds to the assumption that
$p^{\alpha_1\dots \alpha_p}_{i_1\dots \i_p}=0$, $\forall p\geq 2$.)
We have

$$\begin{array}{ccl}
dQ^{i,f} & = & \displaystyle \sum_{\alpha}f^{\alpha}{\partial \over \partial p_i^{\alpha}}\inn \Omega
+ y^i\sum_{\alpha}{\partial f^{\alpha}\over \partial x^{\alpha}}
{\partial \over \partial \epsilon}\inn \Omega,\\
 & & \\
dP_{\mu,g} & = & \displaystyle \sum_{\alpha}{\partial g\over \partial x^{\alpha}}
\Pi^{\alpha}_{\mu}\inn \Omega -
g {\partial \over \partial q^{\mu}}\inn \Omega,
\end{array}$$
where

$$\begin{array}{ccl}
\Pi^{\alpha}_{\beta} & = & \displaystyle 
\delta^{\alpha}_{\beta}\epsilon{\partial \over \partial \epsilon}
+ \sum_{p=1}^{n}{1\over p!^2}\sum_{i_1,\dots ,i_p;\alpha_1,\dots ,\alpha_p}
\left( p^{\alpha_1\dots \alpha_p}_{i_1\dots i_p}\delta^{\alpha}_{\beta}
- \sum_{j=1}^p
p{^{\alpha_1\dots \alpha_{j-1}}_{i_1\dots i_{j-1}}}
{^{\alpha}_{i_j}}
{^{\alpha_{j+1}\dots \alpha_n}_{i_{j+1}\dots i_n}}
\delta^{\alpha_j}_{\beta}\right)

{\partial \over \partial p^{\alpha_1\dots \alpha_p}_{i_1\dots i_p}}\\
\Pi^{\alpha}_{n+i} & = &\displaystyle 
\sum_{p=0}^{n-1}{1\over p!^2}\sum_{i_1,\dots ,i_p;\alpha_1,\dots ,\alpha_p}
p^{\alpha\alpha_1\dots \alpha_p}_{ii_1\dots i_p}
{\partial \over \partial p^{\alpha_1\dots \alpha_p}_{i_1\dots i_p}}.
\end{array}$$
The pataplectic vector fields are

$$\begin{array}{ccl}
\Xi(Q^{i,f}) & = & \displaystyle - \sum_{\alpha}
f^{\alpha}{\partial \over \partial p_i^{\alpha}}
- y^i\sum_{\alpha}{\partial f^{\alpha}\over \partial x^{\alpha}}
{\partial \over \partial \epsilon}\\
\Xi(P_{\mu,g})&  = & \displaystyle g {\partial \over \partial q^{\mu}} -
\sum_{\alpha}{\partial g\over \partial x^{\alpha}} \Pi^{\alpha}_{\mu}.
\end{array}$$
Finally by using Proposition 1, the Poisson $\mathfrak{p}$-brackets will be

$$\begin{array}{ccl}
\{ Q^{i,f},Q^{j,\tilde{f}}\} & = & 0,\\
 & & \\
\{ P_{i,g},P_{j,\tilde{g}}\} & = & \displaystyle
d\left( g\tilde{g}{\partial \over \partial y^j}\inn 
{\partial \over \partial y^i}\inn \theta\right) ,\\
 & & \\
\{ P_{i,g},Q^{j,f}\} & = & \displaystyle
\delta^j_i
\sum_{\alpha}f^{\alpha}g\omega_{\alpha}.
\end{array}$$

\noindent Hence if $g$ and $\tilde{g}$ have compact support, we
obtain that on any submanifold $S$ of dimension $n-1$ without boundary,

$$\int_S\{ Q^{i,f},Q^{j,\tilde{f}}\} = \int_S \{ P_{i,g},P_{j,\tilde{g}}\} = 0
\quad \hbox{and}\quad
\int_S\{ P_{i,g},Q^{j,f}\} = \delta^j_i\int_S\sum_{\alpha}f^{\alpha}g\omega_{\alpha}$$

\subsection{Extension of the $\mathfrak{p}$-bracket to forms of degree less than $n-1$}

The $\mathfrak{p}$-brackets defined above allow us to express the dynamics
of an observable which is in $\mathfrak{P}^{n-1}{\cal M}$. We shall extend
this bracket to some forms in $\Gamma({\cal M}, \Lambda^pT^{\star}{\cal M})$, where
$0\leq p\leq n-1$. Like in the case $p = n-1$, not every $p$-form is admissible
and, as we shall see, the class of such $p$-forms is quite
restricted and is basically composed
of ``position'' observables. However when the Hamiltonian system is degenerate,
due to some gauge symmetry and constraints, 
some ``momentum'' observable can be represented by $p$-forms with $p<n-1$.
An instance of such a situation is the electromagnetic field studied in
Section 5.3.

For $1\leq p\leq n$, we define $\mathfrak{P}^{p-1}{\cal M}$ to be the set of
sections $a$ of $\Gamma({\cal M}, \Lambda^{p-1}T^{\star}{\cal M})$, such that,
for all $1\leq \alpha_1<\dots <\alpha_{n-p}\leq n$, 

$$dx^{\alpha_1}\wedge \dots \wedge dx^{\alpha_{n-p}}\wedge a\in \mathfrak{P}^{n-1}{\cal M}.$$

We introduce anticommuting (Grassmann) variables $\tau_1,\dots, \tau_n$,
which behave under change of coordinates like
${\partial \over \partial x^1},\dots ,{\partial \over \partial x^n}$.
We shall consider functions and forms depending on the variables
$(\tau_{\alpha},x^{\alpha},y^i,p_{\mu_1,\dots,\mu_n})$. Alternatively they can be seen
as functions on the bundle $\Pi T{\cal X}\otimes _{\cal X}{\cal M}$, where $\Pi T{\cal X}$
is a copy of $T{\cal X}$ in which the parity of vectors in the fibers $T_x{\cal X}$
has been reversed. We consider $^s\Gamma({\cal M}, \Lambda^{n-1}T^{\star}{\cal M})$ to be
the set of $(n-1)$-forms on ${\cal M}$ whose coefficients are in the algebra
$\Bbb{R}[\tau_1,\dots,\tau_n]$. More intrinsically,
$^s\Gamma({\cal M}, \Lambda^{n-1}T^{\star}{\cal M})$ can be identified with
${\cal C}^{\infty}(\Pi T{\cal X })\otimes _{{\cal C}^{\infty}({\cal X})}\Gamma({\cal M}, \Lambda^{n-1}T^{\star}{\cal M})$,
meaning that any form $A\in \;^s\Gamma({\cal M}, \Lambda^{n-1}T^{\star}{\cal M})$ is a finite
sum of terms of the form
$\phi(x,\tau)\theta$, where $\phi\in {\cal C}^{\infty}(\Pi T{\cal X })$ and
$\theta\in \Gamma({\cal M}, \Lambda^{n-1}T^{\star}{\cal M})$. Through this identification
we can define $^s\mathfrak{P}^{n-1}{\cal M}$ to be the subset of
$^s\Gamma({\cal M}, \Lambda^{n-1}T^{\star}{\cal M})$ linearly spanned by
$\phi(x,\tau)\theta$, where $\phi\in {\cal C}^{\infty}(\Pi T{\cal X })$ and
$\theta\in \mathfrak{P}^{n-1}{\cal M}$.

Obviously for any $A\in \;^s\mathfrak{P}^{n-1}{\cal M}$, there exists
some vector field $\Xi(A)$ on ${\cal M}$ with coefficients in
$\Bbb{R}[\tau_1,\dots,\tau_n]$ such that $dA = - \Xi(A)\inn \Omega$.
A more geometrical description of $\Xi(A)$ is that it is a section of the
bundle $\Pi T{\cal X}\otimes _{\cal X}T{\cal M}$ over ${\cal M}$.

We embedd each $\mathfrak{P}^{p-1}{\cal M}$ in $^s\mathfrak{P}^{n-1}{\cal M}$
by 

$$\begin{array}{ccc}
\mathfrak{P}^{p-1}{\cal M} & \longrightarrow & ^s\mathfrak{P}^{n-1}{\cal M}\\
a & \longmapsto & ^sa,
\end{array}$$
where the ``superform'' $^sa$ is defined by

$$^sa := \sum_{\alpha_1<\dots <\alpha_{n-p}}\tau_{\alpha_1}\dots \tau_{\alpha_{n-p}}
dx^{\alpha_1}\wedge \dots \wedge dx^{\alpha_{n-p}}\wedge a.$$
Then

$$\Xi(\;^sa) = \sum_{\alpha_1<\dots <\alpha_{n-p}}\tau_{\alpha_1}\dots \tau_{\alpha_{n-p}}
\Xi(dx^{\alpha_1}\wedge \dots \wedge dx^{\alpha_{n-p}}\wedge a).$$

We endow $^s\mathfrak{P}^{n-1}{\cal M}$ with the Poisson $\mathfrak{p}$-sbracket defined by

$$\{ A,B\}_s := (\Xi(A)\wedge \Xi(B))\inn \Omega,$$
where assuming that $A$ and $B$ are homogeneous in $\tau_{\alpha}$ and are given by

$$A = \sum_{\alpha_1<\dots <\alpha_{n-p}}\tau_{\alpha_1}\dots \tau_{\alpha_{n-p}}
A^{\alpha_1\dots \alpha_{n-p}},$$

$$B = \sum_{\beta_1<\dots <\beta_{n-q}}\tau_{\beta_1}\dots \tau_{\beta_{n-q}}
B^{\beta_1\dots \beta_{n-q}},$$

$$\{A,B\}_s = \sum_{\alpha_1<\dots <\alpha_{n-p}}\sum_{\beta_1<\dots <\beta_{n-q}}
\tau_{\alpha_1}\dots \tau_{\alpha_{n-p}}\tau_{\beta_1}\dots \tau_{\beta_{n-q}}
\Xi(B^{\beta_1\dots \beta_{n-q}})\inn \Xi(A^{\alpha_1\dots \alpha_{n-p}})\inn \Omega.$$

Lemma 2 implies immediately that

$$d\{A,B\}_s = - [\xi_A,\xi_B]_s\inn \Omega,$$
where $\xi_A=\Xi(A)$, $\xi_B=\Xi(B)$ and the (super)-Lie bracket $[.,.]_s$ is defined for homogeneous forms $A$ and $B$ by
$[\xi_A,\xi_B]_sf = \xi_A\inn d(\xi_B\inn df) + (-1)^{|A||B|+1}\xi_B\inn d(\xi_A\inn df)$
(here $|A|$ and  $|B|$ are the homogeneity degrees in the variables $\tau_{\alpha}$).
Furthermore, one can deduce easily from Lemma 3 the following relations for all
homogeneous forms $A$, $B$ and $C$ in $^s\mathfrak{P}^{n-1}{\cal M}$,

$$\{A,B\}_s = (-1)^{|A||B|+1}\{B,A\}_s,$$
\noindent  $\displaystyle (-1)^{|A||C|}\{\{A,B\}_s,C\}_s + (-1)^{|B||A|}\{\{B,C\}_s,A\}_s +
(-1)^{|C||B|}\{\{C,A\}_s,B\}_s = $
$$(-1)^{|A||C|}d((\xi_A\wedge \xi_B\wedge \xi_C)\inn \Omega).$$
Hence $^s\mathfrak{P}^{n-1}{\cal M}$ has the structure of a graded Lie algebra modulo
exact terms.

Now suppose that we can prove that for some forms $a\in \mathfrak{P}^{p-1}{\cal M}$
and $b\in \mathfrak{P}^{q-1}{\cal M}$, $\{\;^sa,\; ^sb\}_s$ is equal to some
$^sc$ where $c\in \mathfrak{P}^{p+q-n-1}{\cal M}$, then we could define the
{\em internal} $\mathfrak{p}$-bracket
between $a$ and $b$ by $\{a,b\}:=c$ . This turns actually to be true for
a simple reason: all these brackets vanish by Proposition 2 below
\footnote{excepted of course in the case where
$p=n$ or $q=n$}.
However this fact is no longer true in general in the interesting
case where we have constraints as shown in Section 5.3.

\begin{lemm}
For $1\leq p <n$, $\mathfrak{P}^{p-1}{\cal M}$ coincides with
$\Lambda^{p-1}T^{\star}({\cal X}\times {\cal Y})$. 
\end{lemm}
{\bf Proof} {\em First step:} let $1\leq p<n$ and
$a\in \mathfrak{P}^{p-1}{\cal M}$ and choose any $1\leq \alpha_1<\dots <\alpha_{n-p}\leq n$,
so that
$dx^{\alpha_1}\wedge \dots \wedge dx^{\alpha_{n-p}}\wedge a\in \mathfrak{P}^{n-1}{\cal M}$.
Let us denote
$\xi:= \Xi(dx^{\alpha_1}\wedge \dots \wedge dx^{\alpha_{n-p}}\wedge a)$. Decompose $\xi$:

$$\xi = \sum_{\mu}\xi^{\mu}{\partial \over \partial q^{\mu}} +
\sum_{\mu_1<\dots <\mu_n}\xi_{\mu_1\dots \mu_n}
{\partial \over \partial p_{\mu_1\dots \mu_n}}.$$
Then

$$\begin{array}{ccl}
- \xi\inn \Omega & = & \displaystyle -\sum_{\mu_1<\dots <\mu_n}\xi_{\mu_1\dots \mu_n}
dq^{\mu_1}\wedge \dots \wedge dq^{\mu_n}\\
 & & \displaystyle - \sum_{\nu}\xi^{\nu}\sum_{\alpha=1}^n(-1)^{\alpha}
\sum_{\footnotesize \begin{array}{c}\mu_1<\dots <\mu_n\\\mu_{\alpha} = \nu\end{array}}
dp_{\mu_1\dots \mu_n}\wedge dq^{\mu_1}\wedge \dots \wedge dq^{\mu_{\alpha-1}}
\wedge dq^{\mu_{\alpha+1}}\wedge \dots \wedge dq^{\mu_n}.
\end{array}$$
This expression should be equal to
$(-1)^{n-p-1}dx^{\alpha_1}\wedge \dots \wedge dx^{\alpha_{n-p}}\wedge da$. Note that
for any $1\leq \nu\leq n+k$, there exist $n$ integers $\mu_1<\dots <\mu_n$
such that $\nu\in \{\mu_1,\dots ,\mu_n\}$ but $\alpha_1\not\in \{\mu_1,\dots ,\mu_n\}$.
This forces $\xi^{\nu}=0$. Hence we are left with

$$(-1)^{n-p-1}dx^{\alpha_1}\wedge \dots \wedge dx^{\alpha_{n-p}}\wedge da =
-\sum_{\mu_1<\dots <\mu_n}\xi_{\mu_1\dots \mu_n}
dq^{\mu_1}\wedge \dots \wedge dq^{\mu_n}$$
which implies that $a$ does not depend on the variables $p_{\mu_1\dots \mu_n}$.
Hence $a\in \Lambda^{p-1}T^{\star}({\cal X}\times {\cal Y})$.\\
{\em Second step:} Conversely let $a\in \Lambda^{p-1}T^{\star}({\cal X}\times {\cal Y})$.
Then, for each $1\leq \alpha_1<\dots <\alpha_{n-p}\leq n$,
$dx^{\alpha_1}\wedge \dots \wedge dx^{\alpha_{n-p}}\wedge a$
belongs to $\Lambda^{n-1}T^{\star}({\cal X}\times {\cal Y})$, which is a subset of
$\in \mathfrak{P}^{n-1}{\cal M}$ by Lemma 4. So
$a\in \mathfrak{P}^{p-1}{\cal M}$. \bbox

\begin{prop}{\bf a)}
For $1\leq p,q <n$, $a\in \mathfrak{P}^{p-1}{\cal M}$ and
$b\in \mathfrak{P}^{q-1}{\cal M}$ the $\mathfrak{p}$-sbracket of
$^sa$ with $^sb$ vanishes: $\{\;^sa,\;^sb\}_s = 0$ and hence the
internal $\mathfrak{p}$-bracket
$\{a,b\}$ exists and is equal to 0.\\
{\bf b)} For $1\leq p <n$, $a\in \mathfrak{P}^{n-1}{\cal M}$ and
$b\in \mathfrak{P}^{p-1}{\cal M}$,

\begin{equation}\label{supercrochet-p-n}
\{\;^sa,\;^sb\}_s = -\;^s(db)\Xi(a)(\tau) + \;^s\{a,b\},
\end{equation}
where $\{a,b\}=\Xi(a)\inn db$ is the external $\mathfrak{p}$-bracket and

$$\Xi(a)(\tau) := \sum_{\alpha}dx^{\alpha}(\Xi(a))\tau_{\alpha}.$$
(It is actually a superfunction on $\Pi T{\cal X}$.)
As a consequence, if $dx^{\alpha}(\Xi(a))=0$, $\forall \alpha$, the
internal $\mathfrak{p}$-bracket
$\{a,b\}$ exists and coincides with the external $\mathfrak{p}$-bracket.
\end{prop}
{\bf Proof} Case {\em a)}: by Lemma 5 $a$ and $b$ are in $\Lambda^{p-1}T^{\star}({\cal X}\times {\cal Y})$
and $\Lambda^{q-1}T^{\star}({\cal X}\times {\cal Y})$ respectively, so $^sa$ and $^sb$ are in
${\cal C}^{\infty}(\Pi T{\cal X})\otimes _{{\cal C}^{\infty}({\cal X})}\Lambda^{n-1}T^{\star}({\cal X}\times {\cal Y})$.
Hence their $\mathfrak{p}$-sbracket vanish by Proposition 1.\\
Let us consider the case {\em b)}.
Let us denote $\xi_a  = \Xi(a)$ and write

$$\xi_a = \sum_{\mu}\xi_a^{\mu}{\partial \over \partial q^{\mu}} +
\sum_{\mu_1<\dots <\mu_n}\xi_{a,\mu_1\dots \mu_n}
{\partial \over \partial p_{\mu_1\dots \mu_n}},$$
then

$$\begin{array}{ccl}
\{a,\;^sb\}_s & = & \displaystyle \sum_{\alpha_1<\dots <\alpha_{n-p}}
(-1)^{n-p}\tau_{\alpha_1}\dots \tau_{\alpha_{n-p}} \xi_a\inn
(dx^{\alpha_1}\wedge \dots \wedge dx^{\alpha_{n-p}}\wedge db)\\
 & = & \displaystyle (-1)^{n-p}\sum_{\alpha}\sum_{\alpha_1<\dots <\alpha_{n-p}} \sum_{l=1}^{n-p}
\delta^{\alpha_l}_{\alpha}\xi^{\alpha}_a
\tau_{\alpha_l}\tau_{\alpha_1}\dots \tau_{\alpha_{l-1}}\tau_{\alpha_{l+1}}
\dots \tau_{\alpha_{n-p}} \\
 &  & \displaystyle 
dx^{\alpha_1}\wedge \dots \wedge dx^{\alpha_{l-1}}\wedge dx^{\alpha_{l+1}}\wedge
\dots \wedge dx^{\alpha_{n-p}}\wedge db\\
 & & \displaystyle + \sum_{\alpha_1<\dots <\alpha_{n-p}} 
\tau_{\alpha_1}\dots \tau_{\alpha_{n-p}} 
dx^{\alpha_1}\wedge \dots \wedge dx^{\alpha_{n-p}}\wedge (\xi_a\inn db)\\
 & & \\
 & = & \displaystyle (-1)^{n-p}\sum_{\alpha}
\xi^{\alpha}_a\tau_{\alpha}\sum_{\alpha_1<\dots <\alpha_{n-p-1}}
\tau_{\alpha_1}\dots \tau_{\alpha_{n-p-1}}
dx^{\alpha_1}\wedge \dots \wedge dx^{\alpha_{n-p-1}}\wedge db\\
 & & + \;^s(\xi_a\inn db)\\
 & & \\
 & = & (-1)^{n-p}\sum_{\alpha}\xi^{\alpha}_a\tau_{\alpha}\;^s(db)
+ \;^s\{a,b\}.
\end{array}$$
And the claim is proved.
\bbox

One simple example is the 0-form $y^j$. The associated superform is

$$^sy^j = \sum_{\alpha}y^j(-1)^{\alpha-1}\tau_1\dots \tau_{\alpha-1}
\tau_{\alpha+1}\dots \tau_n{\partial \over\partial x^{\alpha}}\inn \omega.$$
Since

$$d\;^sy^j = \sum_{\alpha}(-1)^{\alpha-1}\tau_1\dots \tau_{\alpha-1}
\tau_{\alpha+1}\dots \tau_n\omega_{\alpha}^j,$$
we have
$$\Xi(\;^sy^j) = \sum_{\alpha}(-1)^{\alpha}\tau_1\dots \tau_{\alpha-1}
\tau_{\alpha+1}\dots \tau_n{\partial \over\partial p^{\alpha}_j}.$$
Let us compute the $\mathfrak{p}$-sbracket with
$P_i={\partial \over \partial y^i}\inn \theta$:

$$\begin{array}{ccl}
\{P_i,\;^sy^j\}_s & = & \displaystyle \sum_{\alpha}(-1)^{\alpha}\tau_1\dots \tau_{\alpha-1}
\tau_{\alpha+1}\dots \tau_n
{\partial \over\partial p^{\alpha}_j}\inn {\partial \over \partial y^i}\inn \Omega\\
 & = & \displaystyle \sum_{\alpha}(-1)^{\alpha-1}\tau_1\dots \tau_{\alpha-1}
\tau_{\alpha+1}\dots \tau_n \delta^j_i
{\partial \over \partial x_{\alpha}}\inn \omega\\
 & =& \;^s\delta^j_i.
\end{array}$$
Thus

$$\{P_i,y^j\} = \delta^j_i.$$

\subsection{Integral of an observable in $\mathfrak{P}^{p-1}{\cal M}$ and
dynamical equations}

Let $\Gamma$ be a submanifold of dimension $n$ on ${\cal M}$ and
let $D$ be be some oriented submanifold with boundary of dimension $p$
($1\leq p\leq n$) included in $\Gamma$. We consider $D_{\Gamma}$, the fiber bundle
over $D$ whose fibers at the point $m\in D$ is the oriented tangent
space to $\Gamma$ at $m$. 

\begin{defi}
Let $a\in \mathfrak{P}^{p-1}{\cal M}$ and
$\psi \in \Gamma({\cal M},\Lambda^nT^{\star}{\cal M})$. We define the
$\mathfrak{p}$-sbracket 

\begin{equation}\label{superbracket}
\begin{array}{ccl}
\{\psi, \;^sa\}_s & := & \displaystyle -\Xi(\;^sa)\inn \psi\\
 & = & \displaystyle -\sum_{\alpha_1<\dots <\alpha_{n-p}}\tau_{\alpha_1}\dots \tau_{\alpha_{n-p}}
\Xi(dx^{\alpha_1}\wedge \dots \wedge dx^{\alpha_{n-p}}\wedge a)
\inn d\psi. 
\end{array}
\end{equation}

We define the integral of $\{\psi, \;^sa\}_s$ over $D_{\Gamma}$ to be

\begin{equation}\label{superintegral}
\int_{D_{\Gamma}} \{\psi, \;^sa\}_s := \int_{D_{\Gamma}}
-\sum_{\alpha_1<\dots <\alpha_{n-p}}(X_{\alpha_1}\wedge \dots \wedge X_{\alpha_{n-p}}\wedge
\Xi(dx^{\alpha_1}\wedge \dots \wedge dx^{\alpha_{n-p}}\wedge a))\inn
d\psi,
\end{equation}
where $X = X_1\wedge \dots \wedge X_n$ is the $n$-vector tangent to
$\Gamma$ at $m$ such that $dx^{\alpha}(X_{\beta}) = \delta^{\alpha}_{\beta}$.
Notice that this definition does not depend on the parametrisation which
is used.
\end{defi}

\begin{theo}
Assume that $\Gamma$ is
the graph of some solution of the Hamilton equations (\ref{hamilton6}). Then for
any $a\in \mathfrak{P}^{p-1}{\cal M}$, 

\begin{equation}\label{superdyna}
\int_{D} da = \int_{D_{\Gamma}}\{{\cal H}\omega, \;^sa\}_s.
\end{equation}
\end{theo}
{\bf Proof} We can always assume that $D$ is the image of some parametrisation

$$\begin{array}{ccl}
\Delta & \longrightarrow & D\\
t & \longmapsto & {\cal U}(x(t)) = (x(t),u(t),p(t)),
\end{array}$$
where $\Delta$ is an open subset of $\Bbb{R}^p$. Then\\

\noindent $\displaystyle \int_{D_{\Gamma}}\{{\cal H}\omega, \;^sa\}_s$

$$\begin{array}{cl}
= & \displaystyle \int_{\Delta}- 
\sum_{\alpha_1<\dots <\alpha_{n-p}}(X_{\alpha_1}\wedge \dots \wedge X_{\alpha_{n-p}}\wedge
\Xi(dx^{\alpha_1}\wedge \dots \wedge dx^{\alpha_{n-p}}\wedge a))\inn \\
 & \displaystyle 
d{\cal H}\wedge \omega \left(
{\partial {\cal U}\circ x\over \partial t^1},\dots ,
{\partial {\cal U}\circ x\over \partial t^p}\right) 
dt^1\wedge \dots \wedge dt^p\\
= & \displaystyle \int_{\Delta}- 
\sum_{\alpha_1<\dots <\alpha_{n-p}}\sum_{\beta_1<\dots <\beta_p}
d{\cal H}\wedge \omega \left(
X_{\alpha_1},\dots ,X_{\alpha_{n-p}},
\Xi(dx^{\alpha_1}\wedge \dots \wedge dx^{\alpha_{n-p}}\wedge a),
X_{\beta_1},\dots ,X_{\beta_p}\right) \\
 & \displaystyle 
\hbox{det}\left( {\partial x^{\beta_1}\over \partial t^1},\dots ,
{\partial x^{\beta_p}\over \partial t^p}\right)
dt^1\wedge \dots \wedge dt^p.
\end{array}$$
Now by using (\ref{hamilton6}),\\

\noindent $\displaystyle \int_{D_{\Gamma}}\{{\cal H}\omega, \;^sa\}_s$

$$\begin{array}{cl}
= & \displaystyle \int_{\Delta} (-1)^{n-p+1}
\sum_{\alpha_1<\dots <\alpha_{n-p}}\sum_{\beta_1<\dots <\beta_p}
\Omega \left(
\Xi(dx^{\alpha_1}\wedge \dots \wedge dx^{\alpha_{n-p}}\wedge a),
X_{\alpha_1},\dots ,X_{\alpha_{n-p}},
X_{\beta_1},\dots ,X_{\beta_p}\right) \\
 & \displaystyle 
\hbox{det}\left( {\partial x^{\beta_1}\over \partial t^1},\dots ,
{\partial x^{\beta_p}\over \partial t^p}\right)
dt^1\wedge \dots \wedge dt^p\\
= & \displaystyle \int_{\Delta} 
\sum_{\alpha_1<\dots <\alpha_{n-p}}\sum_{\beta_1<\dots <\beta_p}
dx^{\alpha_1}\wedge \dots \wedge dx^{\alpha_{n-p}}\wedge da\left(
X_{\alpha_1},\dots ,X_{\alpha_{n-p}},
X_{\beta_1},\dots ,X_{\beta_p}\right) \\
 & \displaystyle 
\hbox{det}\left( {\partial x^{\beta_1}\over \partial t^1},\dots ,
{\partial x^{\beta_p}\over \partial t^p}\right)
dt^1\wedge \dots \wedge dt^p\\
= & \displaystyle \int_{\Delta} 
\sum_{\beta_1<\dots <\beta_p}
da\left(
X_{\beta_1},\dots ,X_{\beta_p}\right) 
\hbox{det}\left( {\partial x^{\beta_1}\over \partial t^1},\dots ,
{\partial x^{\beta_p}\over \partial t^p}\right)
dt^1\wedge \dots \wedge dt^p\\
= & \displaystyle \int_D da.
\end{array}$$
\bbox
There exists however a much simpler concept of bracket between ${\cal H}\omega$ and
observables in $\mathfrak{P}^{p-1}{\cal M}$ which is also suitable for the dynamical
equation in most cases. Namely we call a form $a$ in $\mathfrak{P}^{p-1}{\cal M}$ an
{\em admissible} form if $\forall \alpha_1<\dots <\alpha_{n-p}$,

\begin{equation}\label{admissible}
\begin{array}{c}\displaystyle 
dx^{\alpha}(\Xi(dx^{\alpha_1}\wedge \dots \wedge dx^{\alpha_{n-p}}\wedge a)) = 0,\quad \forall \alpha\\
\hbox{or equivalentely}\quad dx^{\alpha}(\Xi(\;^sa)) = 0,\quad \forall \alpha.
\end{array}
\end{equation}
The reader may wonder the meaning of this definition since, in view of Lemma 5, all forms in 
$\mathfrak{P}^{p-1}{\cal M}$ are admissible for $p<n$. Again the point is that we may
encounter variational problems with gauge symmetry and constraints for which non admissible forms
exist in $\mathfrak{P}^{p-1}{\cal M}$.

\begin{defi}\label{defi4.4}
Assume that $a\in \mathfrak{P}^{p-1}{\cal M}$ satisfies (\ref{admissible}) and let
$\psi\in \Gamma({\cal M},\Lambda^nT^{\star}{\cal M})$.
Then we define the $\mathfrak{p}$-bracket

$$\{\psi, a\} := - \sum_{\alpha_1<\dots <\alpha_{n-p}}
{\partial \over \partial x^{\alpha_1}}\wedge \dots \wedge 
{\partial \over \partial x^{\alpha_{n-p}}}\wedge 
\Xi(dx^{\alpha_1}\wedge \dots \wedge dx^{\alpha_{n-p}}\wedge a)\inn
d\psi.$$
\end{defi}
\begin{lemm}
Let $a\in \mathfrak{P}^{p-1}{\cal M}$ be an admissible form (i. e. such that
(\ref{admissible}) holds) and let $\Gamma$ be a
$n$-dimensional submanifold of ${\cal M}$ which is a graph over ${\cal X}$. Then for any oriented
submanifold $D$ of dimension $p$ included in $\Gamma$,

\begin{equation}\label{demystifie}
\int_{D_{\Gamma}}\{{\cal H}\omega, \;^sa\}_s =
\int_D\{{\cal H}\omega, a\}.
\end{equation}
\end{lemm}
{\bf Proof} We use the same notations as in the proof of Theorem 3. Because of the
condition (\ref{admissible}),

\noindent $\displaystyle \sum_{\alpha_1<\dots <\alpha_{n-p}}
d{\cal H}\wedge \omega \left(
X_{\alpha_1},\dots ,X_{\alpha_{n-p}},
\Xi(dx^{\alpha_1}\wedge \dots \wedge dx^{\alpha_{n-p}}\wedge a),
X_{\beta_1},\dots ,X_{\beta_{n-p}}\right) $

$$\begin{array}{l}
\displaystyle = \sum_{\alpha_1<\dots <\alpha_{n-p}}(-1)^{n-p}
d{\cal H}(\Xi(dx^{\alpha_1}\wedge \dots \wedge dx^{\alpha_{n-p}}\wedge a))
\omega(X_{\alpha_1},\dots ,X_{\alpha_{n-p}},X_{\beta_1},\dots ,X_{\beta_{n-p}})\\
\displaystyle = \sum_{\alpha_1<\dots <\alpha_{n-p}}(-1)^{n-p}
d{\cal H}(\Xi(dx^{\alpha_1}\wedge \dots \wedge dx^{\alpha_{n-p}}\wedge a))
\omega\left( {\partial \over \partial x^{\alpha_1}},\dots ,{\partial \over \partial x^{\alpha_{n-p}}},
X_{\beta_1},\dots ,X_{\beta_{n-p}}\right) \\
\displaystyle = \sum_{\alpha_1<\dots <\alpha_{n-p}}
d{\cal H}\wedge \omega \left(
{\partial \over \partial x^{\alpha_1}},\dots ,{\partial \over \partial x^{\alpha_{n-p}}},
\Xi(dx^{\alpha_1}\wedge \dots \wedge dx^{\alpha_{n-p}}\wedge a),
X_{\beta_1},\dots ,X_{\beta_{n-p}}\right) \\
\displaystyle = - \sum_{\alpha_1<\dots <\alpha_{n-p}}\{{\cal H}\omega, a\}(X_{\beta_1},\dots ,X_{\beta_{n-p}}).
\end{array}$$
This implies the result by summation over $\beta_1<\dots <\beta_{n-p}$ and integration
over $D$. \bbox

\begin{coro}
Let $a\in \mathfrak{P}^{p-1}{\cal M}$ be an admissible form and let $\Gamma$ be a
$n$-dimensional submanifold of ${\cal M}$ which is a graph over ${\cal X}$ of a solution
of the Hamilton equations (\ref{hamilton6}). Then for any oriented
submanifold $D$ of dimension $p$ included in $\Gamma$,

\begin{equation}
\int_D\{{\cal H}\omega, a\} =
\int_Dda.
\end{equation}
\end{coro}
{\bf Examples} The 0-form $y^i$ and the 1-form $y^idy^j$ are
admissible and

$$\{{\cal H}\omega,y^i\} = \sum_{\alpha}{\partial {\cal H}\over \partial p^{\alpha}_i}
dx^{\alpha},$$
$$\{{\cal H}\omega,y^idy^j\} = \sum_{\alpha<\beta}
{\partial {\cal H}\over \partial p^{\alpha\beta}_{ij}}
dx^{\alpha}\wedge dx^{\beta}.$$

\subsection{Noether theorem}

It is natural to relate the Noether theorem
to the pataplectic structure.

Let $\xi$ be a tangent vector field on ${\cal X}\times {\cal Y}$,
$\xi$ will be an infinitesimal symmetry of the variational problem if

$${\cal L}_{\Xi(P_{\xi})}\left( \theta - {\cal H}\omega\right) = 0,$$
since then the integral $\int_{\Gamma}\theta - {\cal H}\omega$ is invariant
under the action of the flow of $\xi$.
Then for any solution $x\longmapsto (U(x),p(x))$, of the Hamilton equations,
the form $P^{\star}_{\xi}$ is closed along the
graph of this solution. This means that if $\Gamma$ is the graph of $(U,p)$, 

$$dP^{\star}_{\xi |\Gamma} = d\left( \xi \inn (\theta - {\cal H}\omega) \right) _{|\Gamma}
= 0.$$
This is a direct consequence of Theorem 2 and of the following calculation.

\begin{lemm}
For any section $\xi$ of $\Gamma({\cal X}\times {\cal Y}, T({\cal X}\times {\cal Y}))$,
we have the relation

\begin{equation}\label{noether}
\{{\cal H}\omega, P_{\xi}\} = 
{\cal L}_{\Xi(P_{\xi})}\left( \theta - {\cal H}\omega\right)
+ d\left( \xi \inn {\cal H}\omega \right) .
\end{equation}
\end{lemm}

{\bf Proof} Using the definition of $\{{\cal H}\omega, P_{\xi}\}$, we have

$$\begin{array}{ccl}
{\cal L}_{\Xi(P_{\xi})}\left( \theta - {\cal H}\omega\right) & = &
\Xi(P_{\xi})\inn (d\theta - d{\cal H}\wedge \omega )
+d\left( \Xi(P_{\xi})\inn (\theta -{\cal H}\omega)\right) \\
 & = & \Xi(P_{\xi})\inn \Omega - \Xi(P_{\xi})\inn d{\cal H}\wedge \omega
 + d\left( \xi\inn \theta -\xi\inn {\cal H}\omega\right) \\
  & = & -dP_{\xi} + \{{\cal H}\omega,P_{\xi}\} +
  d\left( P_{\xi} -\xi\inn {\cal H}\omega\right) ,
\end{array}$$
and the result follows. \bbox
\begin{rema}
It appears that it will be interesting to study solutions of the
Hamilton equations with the constraint ${\cal H} = 0$. This
is possible, because of the freedom left in the Legendre correspondance, thanks to
the parameter $\epsilon$.
The advantage is that then the energy-momentum observables are
described by $P_{a,g}$ which 
belongs to $\mathfrak{P}^{n-1}{\cal M}$.
\end{rema}
\begin{rema}
As a consequence of these observations it is clear that on the submanifold ${\cal H} = 0$, the set
of Noether currents can be identified with $\mathfrak{P}_P^{n-1}{\cal M}$. So we can interpret the results of Proposition
1 concerning $\mathfrak{P}_P^{n-1}{\cal M}$ by saying that the set of Noether currents equipped with
the $\mathfrak{p}$-bracket is a representation modulo exact terms of the Lie algebra of vector
fields on ${\cal X}\times {\cal Y}$ with the Lie bracket. We recover thus various constructions
of brackets on Noether currents (see for instance \cite{Deligne-Freed}).
\end{rema}

\section{Examples}

We present here some examples from the mathematical Physics in order to illustrate
our formalism. We shall see that, by allowing variants of the above theory,
one can find formalisms which are more adapted to some special situations.

\subsection{Interacting scalar fields}
As the simplest example, consider a system of interacting scalar fields
$\{\phi^1,\dots ,\phi^k\}$ on an oriented (pseudo-)Riemannian manifold
$({\cal X},g)$. One should keep in mind that ${\cal X}$ is a four-dimensional
space-time and $g_{\alpha\beta}$ is a Minkowski metric.
These fields can be seen as a map $\phi$ from ${\cal X}$ to
$\Bbb{R}^k$ with its standard Euclidian structure. The metric $g$ on ${\cal X}$
induces a volume form which reads in local coordinates

$$\omega:= g\;dx^1\wedge \dots \wedge dx^n,\quad \hbox{ where }
g:= \sqrt{|\hbox{det}g_{\alpha\beta}(x)|}.$$
Let $V:{\cal X}\longrightarrow \Bbb{R}^k$ be the
interaction potential of the fields, then the Lagrangian density is

$$L(x,\phi,d\phi) := {1\over 2}g^{\alpha\beta}(x)
{\partial \phi^i\over \partial x^{\alpha}}
{\partial \phi_i\over \partial x^{\beta}} - V(\phi(x)).$$
Here $\phi_i=\phi^i$ and we assume that we sum over all repeated indices.
Alternatively one could work with the volume form being $dx^1\wedge \dots \wedge dx^n$
and the Lagrangian density being $gL$, in order to apply directly the
theory constructed in the previous sections. But we shall not choose this approach here
and use a variant which makes clear the covariance of the problem.

We restrict to the Weyl theory, i. e. we work on the submanifold
${\cal M}_{\tiny{\hbox{Weyl}}}$, as in subsection 2.7. So we introduce the
momentum variables $\epsilon$ and $p^{\alpha}_i$ and we start from the
Cartan form

$$\theta = \epsilon\; \omega + p^{\alpha}_id\phi^i\wedge \omega_{\alpha},$$
where $\omega_{\alpha}:= \partial _{\alpha}\inn \omega$. But here $\omega_{\alpha}$
is not closed in general (because $g$ is not constant), so

$$\Omega = d\theta = d\epsilon\wedge \omega + dp^{\alpha}_i\wedge d\phi^i\wedge \omega_{\alpha}
- p^{\alpha}_i{1\over g}{\partial g\over \partial x^{\alpha}}d\phi^i\wedge \omega.$$
The Legendre transform is given by

$$p^{\alpha}_i = {\partial L\over \partial (\partial_{\alpha}\phi^i)} =
g^{\alpha\beta}{\partial \phi^i\over \partial x^{\beta}}
\Longleftrightarrow
{\partial \phi^i\over \partial x^{\alpha}} =
g_{\alpha\beta}p^{\beta}_i,$$
and the Hamiltonian is

$${\cal H}(x,\phi,p) = \epsilon + {1\over 2}g_{\alpha\beta}
p_i^{\alpha}p_i^{\beta} + V(\phi).$$

We use as conjugate variables the 0-forms $\phi^j$ and the $(n-1)$-forms

$$P_{i,f}:= f(x)p_i^{\alpha}\omega_{\alpha}
= f(x){\partial \over \partial \phi^i}\inn \theta\; \in
\mathfrak{P}^{n-1}{\cal M}_{\tiny{\hbox{Weyl}}}.$$
Taking account of the fact that $\omega_{\alpha}$ is not closed, one find

$$\Xi(P_{i,f}) = f{\partial \over \partial \phi^i}
- {\partial f\over \partial x^{\alpha}}p^{\alpha}_i{\partial \over \partial \epsilon}$$
and

$$\{P_{i,f},\phi^j\} = \Xi(P_{i,f})\inn d\phi^j = f\delta^j_i.$$
Also the observables $\phi^j$ are admissible:
$\Xi(\;^s\phi^j) = \sum_{\alpha}{(-1)^{\alpha}\over g}\tau_1\dots
\tau_{\alpha-1}\tau_{\alpha+1}\dots \tau_n{\partial \over \partial p^{\alpha}_j}$,
and according to Definition \ref{defi4.4},

$$
\{{\cal H}\omega,\phi^j\} = 
{\partial {\cal H}\over \partial p^{\alpha}_j}
dx^{\alpha} = g_{\alpha\beta}p^{\beta}_jdx^{\alpha}.$$
Moreover
$$\{{\cal H}\omega,P_{i,f}\} =
- \Xi(P_{i,f})\inn d({\cal H}\omega)= 
\left( -f {\partial V\over \partial \phi^i}
+ {\partial f\over \partial x^{\alpha}}p^{\alpha}_i\right) \omega.
$$
The dynamical equations are that along the graph of a solution,

$$\begin{array}{ccccl}
{\bf d}\phi^i & = & \displaystyle
\{{\cal H}\omega,\phi^i\} & = &\displaystyle
g_{\alpha\beta}p^{\beta}_idx^{\alpha}\\
{\bf d}(fp^{\alpha}_i\omega_{\alpha}) & = & \displaystyle
\{{\cal H}\omega,P_{i,f}\} & = & \displaystyle
\left( -f {\partial V\over \partial \phi^i}
+ {\partial f\over \partial x^{\alpha}}p^{\alpha}_i\right) \omega.
\end{array}$$
The second equation gives

\begin{equation}\label{exa1}
{f\over g}\left( {\partial g\over \partial x^{\alpha}}p^{\alpha}_i
+g {\partial p^{\alpha}_i\over \partial x^{\alpha}}
+ g{\partial V\over \partial \phi^i}\right) = 0,
\end{equation}
while the first relation gives ${\partial \phi^i\over \partial x^{\alpha}}
= g_{\alpha\beta}p^{\beta}_i$. By substitution in (\ref{exa1}) we
find

$${1\over g}{\partial \over \partial x^{\alpha}}\left( 
g\;g^{\alpha\beta}{\partial \phi^i\over \partial x^{\beta}}\right) +
{\partial V\over \partial \phi^i} = 0,$$
the Euler-Lagrange equations of the problem.

\subsection{The conformal string theory}
We consider maps $u$ from a two-dimensional (pseudo-)Riemannian manifold
$({\cal X},g)$ with values in another (pseudo-)Riemannian manifold
$({\cal Y},h)$ of arbitrary dimension. The most general bosonic action for such maps
is ${\cal L}[u]:= \int_{\cal X}L(x,u,du)\omega$ with
$\omega:= g(x) dx^1\wedge dx^2$ and $g(x):= \sqrt{|\hbox{det}g_{\alpha\beta}(x)|}$
as before, and

$$L(x,u,du):= {1\over 2}\left( h_{ij}(u(x))g^{\alpha\beta}(x)
+ b_{ij}(u(x)){\epsilon^{\alpha\beta}\over g(x)}\right)
{\partial u^i\over \partial x^{\alpha}} 
{\partial u^j\over \partial x^{\beta}},$$
where $b:= \sum_{i<j}b_{ij}(y)dy^i\wedge dy^j$ is a given two-form on
${\cal Y}$ and $\epsilon^{12} = -\epsilon^{21} = 1$,
$\epsilon^{11} = \epsilon^{22} =0$. Hence

$${\cal L}[u] = \int_{\cal X}{1\over 2}h_{ij}(u)g^{\alpha\beta}(x)
{\partial u^i\over \partial x^{\alpha}} 
{\partial u^j\over \partial x^{\beta}}\omega +
u^{\star}b.$$
Setting

$$G^{\alpha\beta}_{ij}(x,y):= h_{ij}(y)g^{\alpha\beta}(x)
+ b_{ij}(y){\epsilon^{\alpha\beta}\over g(x)} = G^{\beta\alpha}_{ji}(x,y),$$
we see that $L(x,u,du) = {1\over 2}G^{\alpha\beta}_{ij}(x,u)
{\partial u^i\over \partial x^{\alpha}} 
{\partial u^j\over \partial x^{\beta}}$ and
the Euler-Lagrange equation for this functional is

\begin{equation}\label{string1}
{1\over g}{\partial \over \partial x^{\alpha}}\left( g\;G^{\alpha\beta}_{ij}(x,u(x))
{\partial u^j\over \partial x^{\beta}}\right)
= {\partial G^{\beta\gamma}_{jk}\over \partial y^i}
{\partial u^j\over \partial x^{\beta}}
{\partial u^k\over \partial x^{\gamma}}.
\end{equation}
More covariant formulations exists for the case $b=0$, which correspond
to the harmonic map equation or when the metric on ${\cal X}$ is
Riemannian using conformal coordinates and complex variables
(see \cite{Helein}). The Cartan-Poincar\'e form on ${\cal M}$
is

$$\theta:= \epsilon\; \omega + \sum_{\alpha,i}p^{\alpha}_i\omega^i_{\alpha}
+ \sum_{i<j}p_{ij}\omega^{ij}_{12},$$
(where $\omega^i_1 = g\;dy^i\wedge dx^2$, $\omega^i_2 = g\;dx^1\wedge dy^i$
and $\omega^{ij}_{12} = g\;dy^i\wedge dy^j$). The pataplectic form is

$$\Omega = d\theta = d\epsilon\wedge \omega +
\sum_{\alpha,i}dp^{\alpha}_i\wedge \omega^i_{\alpha}
+ \sum_{i<j}dp_{ij}\wedge \omega^{ij}_{12}
- \sum_{\alpha,i}{p^{\alpha}_i\over g}{\partial g\over \partial x^{\alpha}}
dy^i\wedge \omega
+ \sum_{i<j}\sum_{\alpha}p_{ij}{\partial g\over \partial x^{\alpha}}
dx^{\alpha}\wedge dy^i\wedge dy^j.$$

The Legendre correspondance is generated by the function

$$W(x,u,v,p):= \epsilon + p^{\alpha}_iv^i_{\alpha} +
p_{ij}v^i_1v^j_2 - L(x,u,v) = \epsilon + p^{\alpha}_iv^i_{\alpha} 
- {1\over 2}M^{\alpha\beta}_{ij}(x,y,p)v^i_{\alpha}v^j_{\beta},$$
where we have denoted

$$M^{\alpha\beta}_{ij}(x,y,p):=
h_{ij}(y)g^{\alpha\beta}(x)
+ \left( {b_{ij}(y)\over g(x)}-p_{ij}\right) \epsilon^{\alpha\beta}
= G^{\alpha\beta}_{ij}(x,y) -p_{ij}\epsilon^{\alpha\beta}.$$
This correspondance is given by the relation ${\partial W\over \partial v^i_{\alpha}} = 0$
which gives

\begin{equation}\label{string2}
M^{\alpha\beta}_{ij}(x,y,p)v^j_{\beta} = p_i^{\alpha}.
\end{equation}
Thus, given $(x,y,p)$, finding $(x,y,v,w)$ such that $(x,u,v,w)\leftrightarrow (x,y,p)$
amounts to solving first the linear system (\ref{string2}) for $v$ and then
$w$ is just $W(x,y,v,p)$. This system has a solution in general in the open subset
${\cal O}$ of ${\cal M}$ on which the matrix

$$M = \left( \begin{array}{cc}
h_{ij}(y)g^{11}(x) & h_{ij}(y)g^{12}(x) + {b_{ij}(y)\over g(x)}-p_{ij}\\
h_{ij}(y)g^{21}(x) - {b_{ij}(y)\over g(x)}+ p_{ij} & h_{ij}(y)g^{22}(x)
\end{array}\right)$$
is invertible.
We remark that ${\cal O}$ contains actually the submanifold
${\cal R}:= \{(x,y,p)\in {\cal M}/ g(x)p_{ij} = b_{ij}(y)\}$, so that the Legendre
correspondance induces a diffeomorphism between $T{\cal Y}\otimes T^{\star}{\cal X}$
and ${\cal R}$.\\

We shall need to define on ${\cal O}$ the inverse of $M$, i e.
$K^{ij}_{\alpha\beta}(x,y,p)$ such that

\begin{equation}\label{string3}
K^{ij}_{\alpha\beta}(x,y,p)M^{\beta\gamma}_{jk}(x,y,p) =
\delta^i_k\delta^{\gamma}_{\alpha}.
\end{equation}
Now we can express the solution of (\ref{string2}) by

\begin{equation}\label{string4}
v^i_{\alpha} = K^{ij}_{\alpha\beta}(x,y,p)p_j^{\beta}
\end{equation}
and the Hamiltonian function is

$${\cal H}(x,y,p):= \epsilon + {1\over 2}K^{ij}_{\alpha\beta}(x,y,p)p_i^{\alpha}p_j^{\beta}.$$
We use as conjugate variables the position functions $y^i$ and the momentum
1-forms

$$P_i:= {\partial \over \partial y^i}\inn \theta
= p^{\alpha}_i\omega_{\alpha} + g\; p_{ij}dy^j.$$

The Poisson brackets are obtained as follows. First concerning
$\{{\cal H}\omega, y^i\}$ we compute
$\Xi(\;^sy^i) = {1\over g}\left( \tau_1{\partial \over \partial p^2_i}
- \tau_2{\partial \over \partial p^1_i}\right)$.
Since $dx^{\alpha}(\Xi(\;^sy^i))= 0$, $\forall \alpha$,

$$\begin{array}{ccl}
\{{\cal H}\omega,y^i\} & = & \displaystyle - {1\over g}\left(
{\partial \over \partial x^1}\wedge {\partial \over \partial p^2_i}
- {\partial \over \partial x^2}\wedge {\partial \over \partial p^1_i}\right)
\inn d{\cal H}\wedge \omega\\
 & = & \displaystyle 
{\partial {\cal H}\over \partial p_i^{\alpha}}dx^{\alpha}
=  K^{ij}_{\alpha\beta}p^{\beta}_jdx^{\alpha}.
\end{array}$$
Next we compute $dP_i$:

$$\begin{array}{ccl}
dP_i & = & \displaystyle dp^{\alpha}_i\wedge \omega_{\alpha} + g\;dp_{ij}\wedge dy^j
+ p^{\alpha}_i{\partial g\over \partial x^{\alpha}}{\omega\over g}
+ p_{ij}{\partial g\over \partial x^{\alpha}}dx^{\alpha}\wedge dy^j\\
& = & \displaystyle - {\partial \over \partial y^i}\inn \Omega.
\end{array}$$
Hence

$$\Xi(P_i) = {\partial \over \partial y^i}.$$

Because of $dx^{\alpha}(\Xi(P_i)) =0$, $\forall \alpha$ and of
Proposition 2 we deduce that

$$\begin{array}{ccl}
\{P_i,y^j\} & = & \Xi(P_i)\inn dy^j = \delta^j_i\\
\{{\cal H}\omega,P_i\} & = & \displaystyle -\Xi(P_i)\inn d({\cal H}\omega)
= - {\partial K^{jk}_{\alpha\beta}\over \partial y^i}p^{\alpha}_jp^{\beta}_k\omega.
\end{array}$$
Notice that, because of (\ref{string3}),

$${\partial K^{jk}_{\alpha\beta}\over \partial y^i} =
- K^{jl}_{\alpha\gamma}{\partial M^{\gamma\delta}_{lm}\over \partial y^i}
K^{mk}_{\delta\beta},$$
and thus

$$\{{\cal H}\omega,P_i\} = {\partial M^{\gamma\delta}_{lm}\over \partial y^i} K^{jl}_{\alpha\gamma}K^{mk}_{\delta\beta}
p^{\alpha}_jp^{\beta}_k\omega.$$

The equations of motion are 

\begin{equation}\label{string5}
\begin{array}{ccccl}
{\bf d}y^i & = & \{{\cal H}\omega, y^i\} & = &
K^{ij}_{\alpha\beta}p^{\beta}_jdx^{\alpha}\\
 & & & & \\
{\bf d}P_i & = & \{{\cal H}\omega,P_i\} & = & \displaystyle
{\partial M^{\gamma\delta}_{lm}\over \partial y^i}
K^{jl}_{\alpha\gamma}K^{mk}_{\delta\beta}p^{\alpha}_jp^{\beta}_k\omega,
\end{array}
\end{equation}
along the graph $\Gamma$ of any solution of the Hamilton equations. From the first equation
we deduce that

\begin{equation}\label{string6}
{\partial y^i\over \partial x^{\alpha}} = K^{ij}_{\alpha\beta}p^{\beta}_j
\quad \Longleftrightarrow \quad
p^{\alpha}_i = M^{\alpha\beta}_{ij}{\partial y^j\over \partial x^{\beta}}.
\end{equation}
Now using (\ref{string6}) we see that along $\Gamma$,

$$\begin{array}{ccl}
P_{i|\Gamma} & = & (p^{\alpha}_i\omega_{\alpha} + g\;p_{ij}dy^j)_{|\Gamma}\\
 & = & \displaystyle \left( M^{\alpha\beta}_{ij}{\partial y^j\over \partial x^{\beta}}
+ p_{ij}\epsilon^{\alpha\beta}{\partial y^j\over \partial x^{\beta}}\right) \omega_{\alpha}\\
 & = & \displaystyle G^{\alpha\beta}_{ij}{\partial y^j\over \partial x^{\beta}}
\omega_{\alpha}
= \left( h_{ij}g^{\alpha\beta} + {b_{ij}\over g}\epsilon^{\alpha\beta}\right)
{\partial y^j\over \partial x^{\beta}} \omega_{\alpha},
\end{array}$$
and so the left hand side of the second equation of (\ref{string5}) is

$$dP_{i|\Gamma} = {1\over g}{\partial \over \partial x^{\alpha}}
\left[ g\;G^{\alpha\beta}_{ij}{\partial y^j\over \partial x^{\beta}}\right]
\omega.$$
And still using (\ref{string6}) the right hand side of the second equation of (\ref{string5})
along $\Gamma$ is

$${\partial M^{\alpha\beta}_{jk}\over \partial y^i}
{\partial y^j\over \partial x^{\alpha}}{\partial y^k\over \partial x^{\beta}}\omega.$$
Hence we recover the Euler-Lagrange equation (\ref{string1}).

\subsection{The electromagnetic field}

Here the field is a 1-form $A = A_{\alpha}dx^{\alpha}$ defined on the (pseudo-)Riemannian
manifold ${\cal X}$ (which can also be thougt as a connection 1-form on a
U(1)-bundle). Its differential $dA:= F$ is the electromagnetic field.
We still denote $g_{\alpha\beta}$ the metric on ${\cal X}$
and $\omega = gdx^1\wedge \dots \wedge dx^n$ the volume form.
We are given some vector field $\vec{j} = j^{\alpha}\partial_{\alpha}$ 
on ${\cal X}$ (the electric current field) and we define the
Lagrangian density by

$$L(x,A,dA): = -{1\over 4}F_{\alpha\beta}F^{\alpha\beta} - j^{\alpha}A_{\alpha},$$
where $F_{\alpha\beta}:= \partial_{\alpha}A_{\beta} - \partial_{\beta}A_{\alpha}$
and $F^{\alpha\beta}:= g^{\alpha\gamma}g^{\beta\delta}F_{\gamma\delta}$.

The Euler-Lagrange equation could be written

\begin{equation}\label{max1}
{1\over g}{\partial \over \partial x^{\alpha}}\left( gF^{\alpha\beta}\right) \omega_{\beta}
= j^{\beta}\omega_{\beta},
\end{equation}
or using the notations

$$\star F := \sum_{\alpha<\beta}F^{\alpha\beta}{\partial \over \partial x^{\alpha}}\inn
{\partial \over \partial x^{\beta}}\inn \omega\quad \hbox{and}\quad
j:= j^{\alpha}\omega_{\alpha},$$

\begin{equation}\label{max2}
d(\star F) = j.
\end{equation}
We remark that in order to have solutions it is necessary to suppose that $dj = 0$, which
is the electric charge conservation law\footnote{note also that by writing the
electromagnetic functional $\int {1\over 2}F\wedge \star F - A\wedge j$, one
sees  immediately that the condition $dj=0$ ensures that this functional
is invariant by gauge transformations $A\longmapsto A+df$ up to the addition of
$\int d(fj)$}. 

In our framework the fact that the field is a 1-form means that
we replace the configuration space ${\cal X}\times {\cal Y}$ by
$T^{\star}{\cal X}$. Thus the pataplectic manifold is
${\cal M}:= \Lambda^nT^{\star}(T^{\star}{\cal X})$. We shall here restrict to the
``Weyl''  submanifold of
${\cal M}$ (see Subsection 2.7) which is described by

$${\cal M}_{\tiny{\hbox{Weyl}}}:= 
\{ (x,A,p)/ x\in {\cal X}, A\in T^{\star}_x{\cal X},
p \in \Lambda^nT^{\star}_{(x,A)}(T^{\star}{\cal X}),
\partial_{A_{\alpha}}\wedge \partial_{A_{\beta}}\inn p = 0,\forall \alpha, \beta \}.$$

The latter condition on $p$ means that in local coordinates,
$p = \epsilon\; \omega + \sum_{\alpha,\beta}p^{A_{\alpha}\beta}
dA_{\alpha}\wedge \omega_{\beta}$.
The Cartan form is

$$\theta:= \epsilon\; \omega + \sum_{\alpha,\beta}p^{A_{\alpha}\beta}
dA_{\alpha}\wedge \omega_{\beta},$$
with still $\omega_{\beta}:= \partial_{\beta}\inn \omega$ and the
pataplectic form is

$$\Omega:= d\theta = d\epsilon\wedge \omega + \sum_{\alpha,\beta}dp^{A_{\alpha}\beta}\wedge
dA_{\alpha}\wedge \omega_{\beta} - \sum_{\alpha,\beta}p^{A_{\alpha}\beta}
{1\over g}{\partial g\over \partial x^{\beta}}dA_{\alpha}\wedge \omega.$$
Computing the Legendre transform in ${\cal M}_{\tiny{\hbox{Weyl}}}$,
using $W(x,A,dA,p) = \epsilon + \sum_{\alpha,\beta}p^{A_{\alpha}\beta}\partial_{\beta}A_{\alpha} - L(x,A,dA)$
gives the momenta 

$$p^{A_{\alpha}\beta}:= {\partial L\over \partial (\partial_{\beta}A_{\alpha})}=
F^{\alpha\beta}.$$
We see that the Legendre transform works only provided the compatibility condition

\begin{equation}\label{constraint}
p^{A_{\alpha}\beta} + p^{A_{\beta}\alpha} = 0
\end{equation}
is satisfied. It is an example of a Dirac primary constraint.
Henceforth we shall be restricted to the submanifold 

$${\cal M}_{\tiny{\hbox{Maxwell}}}:= 
\{ (x,A,p)\in {\cal M}_{\tiny{\hbox{Weyl}}}/
p^{A_{\alpha}\beta} + p^{A_{\beta}\alpha} = 0 \},$$
along which we are able to obtain an expression for the
Hamiltonian
$${\cal H}(x,A,p)  = \epsilon -{1\over 4}g_{\alpha\gamma}g_{\beta\delta}
p^{A_{\alpha}\beta}p^{A_{\gamma}\delta} + j^{\alpha}A_{\alpha}.$$
A naive use of this Hamiltonian function 
leads to incorrect dynamical equations. Another possibility,
which was already proposed in \cite{kanatchikov1} is to use as dynamical
variable the 1-form on ${\cal M}_{\tiny{\hbox{Maxwell}}}$

$$ A:=A_{\alpha}dx^{\alpha}.$$
Note that here $A_{\alpha}$ is not a local function of $x$ but an
independant variable. 
Then, as it will be proved below, the momentum variable canonically conjugate to $A$ 
may be chosen to be the $(n-2)$-form

$$\pi:= {1\over 2}\sum_{\alpha,\beta}p^{A_{\alpha}\beta} {\partial \over \partial x^{\alpha}}\inn
{\partial \over \partial x^{\beta}}\inn \omega.$$
Its associated superform is
$$\begin{array}{ccl}
^s\pi & = & \displaystyle 
{1\over 2}\sum_{\alpha,\beta}p^{A_{\alpha}\beta}
\left( \tau_{\alpha}{\partial \over \partial x^{\beta}}
-  \tau_{\beta}{\partial \over \partial x^{\alpha}}\right) \inn \omega\\
 & = & \displaystyle 
\sum_{\alpha,\beta}p^{A_{\alpha}\beta}
\tau_{\alpha}{\partial \over \partial x^{\beta}}\inn \omega.
\end{array}$$
Hence using (\ref{constraint}) 
$$\begin{array}{ccl}
d\;^s\pi & = & \displaystyle 
\sum_{\alpha,\beta}\tau_{\alpha}dp^{A_{\alpha}\beta}
\wedge \left( {\partial \over \partial x^{\beta}}\inn \omega\right)
+ \sum_{\alpha,\beta}{p^{A_{\alpha}\beta}\over g}
\tau_{\alpha}{\partial g\over \partial x^{\beta}}\omega\\
 & = & \displaystyle - \sum_{\alpha}\tau_{\alpha}
{\partial \over \partial A_{\alpha}}\inn \Omega
\end{array}$$
and

$$\Xi(\;^s\pi) = \sum_{\alpha}\tau_{\alpha}
{\partial \over \partial A_{\alpha}}.$$

We also have (denoting $\omega_{\beta}:=
{\partial \over \partial x^{\beta}}\inn \omega$ and
${\tau_1^{\dots}}{^{\alpha}_{\dots}}{^{\dots}_{\dots}}{^{\beta}_{\dots}}{_n^{\dots}}:=
\tau_1\dots \tau_{\alpha-1}\tau_{\alpha+1}\dots \tau_{\beta-1}\tau_{\beta+1}\dots \tau_n$)

$$ ^sA = 
\sum_{\alpha<\beta} (-1)^{n+\alpha+\beta} {A_{\alpha}\over g}
{\tau_1^{\dots}}{^{\alpha}_{\dots}}{^{\dots}_{\dots}}{^{\beta}_{\dots}}{_n^{\dots}} \;\omega_{\beta} 
- \sum_{\beta<\alpha} (-1)^{n+\alpha+\beta} {A_{\alpha}\over g}
{\tau_1^{\dots}}{^{\beta}_{\dots}}{^{\dots}_{\dots}}{^{\alpha}_{\dots}}{_n^{\dots}} \;\omega_{\beta}$$

and

$$\Xi(\;^sA) = \sum_{\alpha<\beta} (-1)^{n+\alpha+\beta+1} {A_{\alpha}\over g}
{\tau_1^{\dots}}{^{\alpha}_{\dots}}{^{\dots}_{\dots}}{^{\beta}_{\dots}}{_n^{\dots}}
{\partial \over \partial p^{A_{\alpha}\beta}}
- \sum_{\beta<\alpha} (-1)^{n+\alpha+\beta+1} {A_{\alpha}\over g}
{\tau_1^{\dots}}{^{\beta}_{\dots}}{^{\dots}_{\dots}}{^{\alpha}_{\dots}}{_n^{\dots}}
{\partial \over \partial p^{A_{\alpha}\beta}}.$$

A computation using (\ref{constraint}) gives

$$\begin{array}{ccl}
\{\;^s\pi,\;^sA\}_s & = & \displaystyle
\sum_{\alpha,\beta,\gamma} (-1)^{n+\alpha+\beta+1} \tau_{\gamma}\left( 
\sum_{\alpha<\beta} {A_{\alpha}\over g}
{\tau_1^{\dots}}{^{\alpha}_{\dots}}{^{\dots}_{\dots}}{^{\beta}_{\dots}}{_n^{\dots}}
- \sum_{\beta<\alpha} {A_{\alpha}\over g}
{\tau_1^{\dots}}{^{\beta}_{\dots}}{^{\dots}_{\dots}}{^{\alpha}_{\dots}}{_n^{\dots}}
\right) {\partial \over \partial p^{A_{\alpha}\beta}}
\inn {\partial \over \partial A_{\gamma}}\inn \Omega\\
 & = & \displaystyle
2(-1)^n(n-1)\sum_{\alpha}\tau_1\dots \tau_{\alpha-1}\tau_{\alpha+1}\dots \tau_n 
{\partial \over \partial x^{\alpha}}\inn dx^1\wedge \dots \wedge dx^n\\
 & = & ^s(2(-1)^n(n-1)).
\end{array}$$
Thus

$$\{\pi,A\} = 2(-1)^n(n-1).$$

As it may be anticipated by Corollary 1, the dynamical equations are described
by the following identities to be true along the graph of a solution of the
Hamilton equations

$$\begin{array}{ccl}
{\bf d}A  & = & \{{\cal H}\omega,A\} \\
& = & (-1)^{n+\alpha+\beta}
\left[ \sum_{\alpha<\beta}{1\over g}
{\partial \over \partial x^1}\wedge \dots
{\partial \over \partial x^{\alpha-1}}\wedge 
{\partial \over \partial x^{\alpha+1}}\wedge \dots 
{\partial \over \partial x^{\beta-1}}\wedge 
{\partial \over \partial x^{\beta+1}}\wedge \dots 
{\partial \over \partial x^n}\wedge 
{\partial \over \partial p^{A_{\alpha}\beta}}\right.\\
 &  & - \left. \sum_{\beta<\alpha}{1\over g}
{\partial \over \partial x^1}\wedge \dots
{\partial \over \partial x^{\beta-1}}\wedge 
{\partial \over \partial x^{\beta+1}}\wedge \dots 
{\partial \over \partial x^{\alpha-1}}\wedge 
{\partial \over \partial x^{\alpha+1}}\wedge \dots 
{\partial \over \partial x^n}\wedge 
{\partial \over \partial p^{A_{\alpha}\beta}}\right]
\inn d{\cal H}\wedge \omega\\
 & = & \displaystyle
\sum_{\alpha,\beta}
{\partial {\cal H}\over \partial p^{A_{\alpha}\beta}}
dx^{\beta}\wedge dx^{\alpha}\\
 & = & \displaystyle
\sum_{\alpha<\beta}
g_{\alpha\gamma}g_{\beta\delta}p^{A_{\gamma}\delta}dx^{\alpha}\wedge dx^{\beta},
\end{array}$$

$$\begin{array}{ccl}
{\bf d}\pi  & = & \displaystyle
\{{\cal H}\omega,\pi\} = -\sum_{\alpha}{\partial \over \partial x^{\alpha}}\wedge 
{\partial \over \partial A_{\alpha}}\inn d{\cal H}\wedge \omega\\
 & = & \displaystyle
{\partial {\cal H}\over \partial x^{\alpha}}\omega_{\alpha}
= j^{\alpha}\omega_{\alpha} = j.
\end{array}$$

The first equation leads to
${\partial A_{\beta}\over \partial x^{\alpha}} - {\partial A_{\alpha}\over \partial x^{\beta}}
= g_{\alpha\gamma}g_{\beta\delta}p^{A_{\gamma}\delta}$ which implies
$F_{\alpha\beta} = g_{\alpha\gamma}g_{\beta\delta}p^{A_{\gamma}\delta}$ or equivalentely
$p^{A_{\alpha}\beta} = F^{\alpha\beta}$. This can be translated into the
relation

$$\pi = \star F,\quad \hbox{along }\Gamma.$$
By substitution in the second equation, {\bf d}$\pi = j$, it gives immediately
(\ref{max2}).\\

A last observation is that infinitesimal gauge transformations $\delta A = df$
(for $f\in {\cal C}^{\infty}({\cal X})$) are generated by the Poisson bracket with
$df\wedge \pi$. We have indeed

$$d(df\wedge \pi) = -df\wedge d\pi = - \sum_{\alpha}{\partial f\over \partial x^{\alpha}}
{\partial \over \partial A_{\alpha}}\inn \Omega,$$
so that $df\wedge \pi \in \mathfrak{P}^{n-1}{\cal M}$ and
$\Xi(df\wedge \pi) = \sum_{\alpha}{\partial f\over \partial x^{\alpha}}
{\partial \over \partial A_{\alpha}}$.
We deduce that

$$\begin{array}{ccl}
\{df\wedge \pi,\pi\} & = & \Xi(df\wedge \pi)\inn d\pi = 0\\
\{df\wedge \pi,A\} & = & \Xi(df\wedge \pi)\inn dA = df.
\end{array}$$
Notice that we could replace $df\wedge \pi$ by $-fd\pi$ or $f(j-d\pi)$ without
changing the brackets with $\pi$ and $A$.

\section{Conclusion}

We obtained an Hamiltonian formulation for variational problems
with an arbitrary number of variables. This could be the starting point for
building a fully relativistic quantum field theory
without requiring the space-time to be Minkowskian. This will be the subject of a
forthcoming paper.
Notice also that we may enlarge the concept of pataplectic manifolds as
manifolds equipped with a closed $n+1$-form and extend to this context
notions like the $\mathfrak{p}$-bracket.\\\\


\begin{thebibliography}{99}

\newcommand{\bib}[1]{\bibitem{#1}}
\bib{De Donder} Th. De Donder, {\sl Th\'eorie Invariante du Calcul des
Variations}, Nuov. \'{e}d. (Gauthier-Villars, Paris, 1935)

\bib{Weyl1} H. Weyl, 
{\em Geodesic fields in the calculus of variations, } 
Ann. Math. (2)  36 (1935) 607-629 

\bib{Caratha1} C. Carath\'{e}odory, 
{\em \"Uber die Extremalen und geod\"atischen Felder in der 
Variationsrechnung der mehrfachen Integrale, } 
Acta Sci. Math. (Szeged) 4 (1929) 193-216

\bib{Rund} H. Rund, {\sl The Hamilton-Jacobi Theory in the Calculus of 
Variations}, (D. van Nostrand Co. Ltd., Toronto, etc. 1966) 
(Revised and augmented reprint, Krieger Publ., New York, 1973) 

\bib{Kastrup1} H. Kastrup, 
{\em Canonical theories of Lagrangian dynamical systems in physics, }
Phys. Rep. 101 (1983) 1-167 

\bib{binz} E. Binz, J. \'{S}niatycki and  H. Fisher, 
{\sl Geometry of Classical Fields, } (North-Holland, Amsterdam, 1989) 


\bib{Gotay1} M.J. Gotay,
{\em An exterior differential systems approach to the Cartan form, }  
 in {\sl Symplectic Geometry and Mathematical Physics, } 
eds. P. Donato, C. Duval, e.a. (Birkh\"{a}user, 
Boston, 1991) p. 160-188

\bib{Gotay2} M.J. Gotay, 
{\em A multisymplectic framework for classical field theory 
and the calculus of variations I. Covariant Hamiltonian formalism, }
in {\sl Mechanics. Analysis and Geometry: 
200 Years after Lagrange}, ed. M. Francaviglia (North Holland, 
Amsterdam, 1991) p. 203-235

\bib{Gotay3} M.J. Gotay, 
{\em A multisymplectic framework for classical field theory 
and the calculus of variations II. Space + time decomposition, } 
Diff. Geom. and its Appl. 1 (1991) 375-390

\bib{Hrabak1} S. P. Hrabak
{\em On a multisymplectic formulation of the classical BRST symmetry for first order field
theories PartII: Geometric Structure}, arXiv: math-ph/9901013

\bib{Hrabak2}
S. P. Hrabak {\em On a multisymplectic formulation of the classical BRST symmetry for first
order field theories PartII: Geometric Structure}, arXiv: math-ph/9901012

\bib{kanatchikov1}
I. V. Kanatchikov {\em Canonical structure of classical field theory in the polymomentum phase space},
arXiv:hep-th/9709229

\bib{Kanatchikov2}
I. V. Kanatchikov, {\em On the canonical structure of the De Donder-Weyl covariant Hamiltonian formulation of field theory I.
Graded Poisson brakets and the equation of motion}, hep-th/9312162
\bib{Gotay4} M.J. Gotay, J. Isenberg, J. Marsden and R. Montgomery, 
{\em Momentum  Maps and Classical Relativistic Fields}, arXiv:physics/9801019 
 
\bib{Enriquez1}
A.Echeverria-Enriquez, M.Munoz-Lecanda, N.Roman-Roy, {\em Geometry of multisymplectic Hamiltonian first-order field theories },
math-ph/0004005

\bib{Paufler}
C.Paufler, {\em A vertical exterior derivative in multisymplectic geometry and a graded Poisson bracket for nontrivial geometries},
math-ph/0002032

\bib{Enriquez2}
A.Echeverria-Enriquez, M.Munoz-Lecanda, N.Roman-Roy, {\em On the
multimomentum bundles and the Legendre maps in field theories}, Rep. Math.
Phys., v. 45 (2000) 85, math-ph/9904007
       
\bib{Enriquez3}
A.Echeverria-Enriquez, M.Munoz-Lecanda, N.Roman-Roy, {\em Multivector
field formulation of Hamiltonian field theories}, J.Phys.A, v.32 (1999)
8461, math-ph/9907007

\bib{Sardanashvily5}
 G.Sardanashvily, {\em SUSY-extended field theory}, hep-th/9911108 (appear
in Int.J.Mod.Phys. A (2000))
  
\bib{Giachetta}      
G.Giachetta, L.Mangiarotti, G.Sardanashvily, {\em Covariant Hamilton
equations for field theory}, J.Phys.A, v.32 (1999) 6629, hep-th/9904062
     
\bib{Sardanashvily1}
  G.Sardanashvily, {\em Generalized Hamiltonian Formalism for Field Theory}, ed.
World Scientific, Singapore, 1995

\bib{Sardanashvily2}
L.Mangiarotti and G.Sardanashvily, {\it{Gauge Mechanics}}, ed. World
Scientific, Singapore, 1998

\bib{Sardanashvily3}
 G.Giachetta, L.Mangiarotti and G.Sardanashvily, {\em New Lagrangian and
Hamiltonian Methods in Field Theory}, ed. World Scientific, Singapore, 1997
      
\bib{Rund2} H. Rund, {\em A Cartan form for the field theory of 
Carath\'eodory in the calculus of variations of multiple integrals}, 
in {\sl Differential Geometry, Calculus of Variations and Their 
Applications}, Lect. Notes Pure and Appl. Math. vol. 100, 
ed. G.M. Rassias and T.M. Rassias, (Marcel Dekker etc., 1985) 
p. 455-469 

\bib{Martin} D. H. Martin,
{\em Canonical variables and geodesic fields for the calculus of
variations of multiple integrals in parametric form}, Math. Z. 104 (1968), 16-27.

\bib{Goldschmidt1} H. Goldschmidt and S. Sternberg, 
{\em The Hamilton-Cartan formalism in the calculus of variations, } Ann. Inst. Fourier (Grenoble)23, fasc. 1 (1973) 203-267.

\bib{Kijowski1} J. Kijowski, 
{\em A finite dimensional canonical formalism in the classical field theory, }
Comm. Math. Phys. 30 (1973) 99-128;\\ 
J. Kijowski and W. Szczyrba, 
{\em A Canonical Structure for Classical Field Theories, } Comm. Math. Phys. 46 (1976) 183-206  

\bib{Helein} F. H\'elein,
{\em Probl\`emes variationnels invariants par transformation conforme en
dimension 2}, to appear in {\it{Partial differential equations and variational calculus in Physics}}, ed. Joseph Kouneiher.

\bib{Gawcedzki1} K. Gaw\c edzki, 
{\em On the generalization of the canonical formalism in the 
classical field theory, }
Rep. Math. Phys. 3 (1972) 307-326; 

\bib{Dedecker1} P. Dedecker, 
{\em Calcul des variations, formes diff\'erentielles et champs 
g\'eod\'esiques, } 
in {\it G\'eometrie Differentielle, } Colloq. Intern. du CNRS LII, 
Strasbourg 1953, (Publ. du CNRS, Paris, 1953) p. 17-34

\bib{Dedecker2} P. Dedecker, 
{\em On the generalization of symplectic geometry to multiple integrals in the calculus of variations, }
in {\it Differential Geometrical Methods in Mathematical Physics,} 
eds. K. Bleuler and A. Reetz, Lect. Notes Math. vol. 570 
(Springer-Verlag, Berlin etc., 1977) p. 395-456

\bib{Deligne-Freed} P. Deligne, D. Freed,
{\em Classical field theory}, in {\it Quantum fields and strings: a
course for mathematicians, Volume 1}, P. Deligne, P. Etingof, D.S. Freed, L.C. Jeffrey,
D. Kahzdan, J.W. Morgan, D.R. Morrison and E. Witten, editors, American Mathematical
Society, 1999.



\end{thebibliography}
\end{document}